\documentclass[conference,compsoc]{IEEEtran}
\usepackage[T1]{fontenc}
\usepackage{newunicodechar}
\newunicodechar{’}{\textquotesingle}
\usepackage{amsmath,amssymb,amsthm}
\usepackage{graphicx,booktabs,array,multirow}
\usepackage{algorithm}
\usepackage{algpseudocode}
\usepackage[table]{xcolor}
\usepackage{etoolbox}
\usepackage{url}
\usepackage[most]{tcolorbox}
\newtcolorbox{answerBox}[1][]{
    colback=gray!8,
    colframe=gray!15,
    boxrule=0.2mm,
    arc=0.5mm,
    width=\columnwidth,
    left=1mm,
    right=1mm,
    top=0.5mm,
    bottom=0.5mm,
    #1}
\definecolor{TableHead}{HTML}{FFFFFF}
\definecolor{TableBlue}{HTML}{FFFFFF}
\definecolor{Heat0}{HTML}{00B887}
\definecolor{Heat1}{HTML}{0DBC8D}
\definecolor{Heat2}{HTML}{1ABF93}
\definecolor{Heat3}{HTML}{26C399}
\definecolor{Heat4}{HTML}{33C69F}
\definecolor{Heat5}{HTML}{47CCA9}
\definecolor{Heat6}{HTML}{FF898D}
\definecolor{Heat7}{HTML}{FF7C81}
\definecolor{Heat8}{HTML}{FF7479}
\definecolor{Heat9}{HTML}{FF6B71}
\definecolor{Heat10}{HTML}{FF6369}
\definecolor{Heat11}{HTML}{FF5B61}
\definecolor{RecoveryDot}{HTML}{FF5B61}
\definecolor{ControlDot}{HTML}{00B887}
\newcommand{\TableHeaderStrut}{\rule[-3.5pt]{0pt}{12.5pt}}
\newcommand{\TableTop}{\hline\hline}
\newcommand{\TableHeadRule}{\hline}
\newcommand{\TableBottom}{\hline\hline}
\AtBeginEnvironment{table}{\renewcommand{\arraystretch}{1.15}\setlength{\extrarowheight}{0.2pt}}
\AtBeginEnvironment{table*}{\renewcommand{\arraystretch}{1.15}\setlength{\extrarowheight}{0.2pt}}
\newcommand{\MetricShade}[2]{%
  \ifdim #1<0.01pt\cellcolor{Heat0}%
  \else\ifdim #1<0.1pt\cellcolor{Heat1}%
  \else\ifdim #1<1pt\cellcolor{Heat2}%
  \else\ifdim #1<5pt\cellcolor{Heat3}%
  \else\ifdim #1<25pt\cellcolor{Heat4}%
  \else\ifdim #1<50pt\cellcolor{Heat5}%
  \else\ifdim #1<75pt\cellcolor{Heat6}%
  \else\ifdim #1<95pt\cellcolor{Heat7}%
  \else\ifdim #1<99pt\cellcolor{Heat8}%
  \else\ifdim #1<99.9pt\cellcolor{Heat9}%
  \else\ifdim #1<99.99pt\cellcolor{Heat10}%
  \else\cellcolor{Heat11}\fi\fi\fi\fi\fi\fi\fi\fi\fi\fi\fi #2}
\newcommand{\UnitMetricShade}[2]{%
  \ifdim #1<0.0001pt\cellcolor{Heat0}%
  \else\ifdim #1<0.001pt\cellcolor{Heat1}%
  \else\ifdim #1<0.01pt\cellcolor{Heat2}%
  \else\ifdim #1<0.05pt\cellcolor{Heat3}%
  \else\ifdim #1<0.25pt\cellcolor{Heat4}%
  \else\ifdim #1<0.5pt\cellcolor{Heat5}%
  \else\ifdim #1<0.75pt\cellcolor{Heat6}%
  \else\ifdim #1<0.95pt\cellcolor{Heat7}%
  \else\ifdim #1<0.99pt\cellcolor{Heat8}%
  \else\ifdim #1<0.999pt\cellcolor{Heat9}%
  \else\ifdim #1<0.9999pt\cellcolor{Heat10}%
  \else\cellcolor{Heat11}\fi\fi\fi\fi\fi\fi\fi\fi\fi\fi\fi #2}
\newcommand{\RateCell}[1]{\MetricShade{#1pt}{#1}}
\newcommand{\RateShade}[2]{\MetricShade{#1pt}{#2}}

\definecolor{CitationPurple}{HTML}{76529A}
\definecolor{ReferenceWine}{HTML}{9B3A55}
\usepackage[colorlinks=true,citecolor=CitationPurple,linkcolor=ReferenceWine,
  urlcolor=CitationPurple,pdfauthor={Chengzhi Yi, Xiao Cheng, Wei Song, Haoyu Wang},pdftitle={When Shuffling Fails: Token Recovery from Obfuscated KV Caches}]{hyperref}
\DeclareUrlCommand\path{\urlstyle{rm}}
\graphicspath{{figures/}}
\newtheorem{theorem}{Theorem}
\newtheorem{proposition}[theorem]{Proposition}

\newcommand{\ones}{\mathbf{1}}
\newcommand{\R}{\mathbb{R}}
\newcommand{\T}{\mathsf{T}}
\newcommand{\scheme}{KV-Cloak}
\newcommand{\veilkv}{\textsc{VeilKV}}

\newcommand{\Stab}{\operatorname{Stab}}
\newcommand{\argmax}{\operatorname*{arg\,max}}
\begin{document}
\bstctlcite{bibliographycontrol}
\title{When Shuffling Fails:\\Token Recovery from Obfuscated KV Caches}
\author{
\IEEEauthorblockN{Chengzhi Yi\IEEEauthorrefmark{1}, Xiao Cheng\IEEEauthorrefmark{2}, Wei Song\IEEEauthorrefmark{3}, Haoyu Wang\IEEEauthorrefmark{1}}
\IEEEauthorblockA{\IEEEauthorrefmark{1}Huazhong University of Science and Technology\\
\IEEEauthorrefmark{2}Macquarie University \qquad \IEEEauthorrefmark{3}Griffith University\\
\texttt{yimou@hust.edu.cn}, \texttt{jumormt@gmail.com}, \texttt{w.song@griffith.edu.au}, \texttt{haoyuwang@hust.edu.cn}}
}
\maketitle
\raggedbottom
\begin{abstract}
LLM serving systems routinely store, offload, and share KV cache blocks for efficiency, exposing token-level representations to infrastructure operators and co-tenants. A class of recent defenses, which we term shuffled obfuscation, protects these blocks by combining secret linear transforms with fresh random permutations, claiming security from the $(b!)$ size of the permutation group. We identify input symmetry as a fundamental flaw in this security argument. The $(b!)$ count measures how many shufflings the defender can sample, but security depends on how many produce distinguishable outputs under the attacker\textquotesingle{}s chosen input. A repeated-token prompt produces identical Value-layer rows across an entire cache block, so every row permutation acts as the identity on the data matrix, and only auxiliary structures such as a single-row recovery marker still vary. The effective state space therefore collapses from $(b!)$ to as few as $(b)$.

Building on this observation, we construct the Repeat-to-Reveal attack. The attacker collects $(b)$ distinguishable protected states from its own probe blocks and exploits their rank-one algebraic differences to recover an equivalent denoiser, without access to the defender’s secret parameters. Crucially, this denoiser transfers across requests within the same secret epoch, so a one-time calibration enables passive token-multiset recovery from any subsequently observed cache block through public-model dictionary matching. We instantiate the attack on \scheme{} and evaluate it on 3{,}000 real user prompts from WildChat across nine models and eleven configurations. The attack near-perfectly recovers the defense\textquotesingle{}s secret parameters, achieving 100\% token-multiset recovery on GPU. Refreshing the row-level secrets between epochs interrupts cross-request transfer, confirming that secret reuse scope is the operative vulnerability. The orbit collapse is further confirmed on two other open-source defenses in the same paradigm, the sequence-permutation implementation of Thomas et al.@ and PermLLM. We were invited by the developers of \scheme{} to submit a detailed report through their private vulnerability reporting channel.
\end{abstract}

\section{Introduction}
KV caches store the Key and Value vectors produced during language-model inference so that later tokens can attend to earlier context without recomputation. Serving systems routinely store, offload, and share cached states across requests and devices for efficiency~\cite{paged}. These operations expose a privacy surface: the cached vectors encode token-level representations tied to the user's input, and any party that reads the stored or transferred state observes content-bearing data even when the client connection itself is protected.

Recent defenses conceal inference representations using secret transforms and permutations. Examples include permutation-based private inference~\cite{permllm,hidden} and KV cache protection~\cite{kvcloak}. Related approaches use refreshed batch mixing~\cite{gelo}, orthogonal conjugation~\cite{conjformer}, or weight and activation masking in partial-TEE inference~\cite{collapse,partialtee}. We call designs combining secret linear transforms with fresh permutations \emph{shuffled obfuscation}. In permutation-based designs such as PermLLM and KV-Cloak, the security argument credits the permutation group's size: $b!$ possible shufflings of $b$ elements appear to make brute-force recovery infeasible~\cite{permllm,kvcloak}.

We show that this $b!$ barrier is illusory. The permutation count measures how many shufflings the defender can sample, but \emph{security depends on how many of them produce distinguishable outputs on the attacker's chosen input}. When that input has symmetry under the permutation group, multiple shufflings act identically, and the observable orbit collapses. In the extreme case, a repeated-token input produces identical cache rows: every row permutation leaves the data matrix unchanged, and only auxiliary structures such as recovery markers still vary. The effective state space can shrink from $b!$ to as few as $b$. Within this reduced orbit, reused structural secrets provide fixed reference points that enable algebraic identification of the defense parameters. The true security measure is not the group order but the orbit size under adversarially chosen inputs.

We demonstrate the attack on \scheme{}~\cite{kvcloak}, a representative state-of-the-art instance of this shuffled obfuscation defense paradigm. \scheme{} transforms each stored KV cache block through four operations: a secret orthogonal column rotation, a single-row additive marker for legitimate decoding, a fresh random row permutation, and a secret orthogonal row transform. The column rotation can be folded into model weights to eliminate its online cost. Among the four operations, only the row permutation is freshly sampled per block. The column rotation, row transform, and marker are reused across requests within a secret epoch. The defense credits this $b!$-sized per-block randomization as the primary barrier against content recovery.

We exploit \emph{input symmetry} to turn this reuse into a calibration signal. Because the other three components persist across requests, the attacker can craft an input that renders the fresh permutation ineffective. A legal prompt consisting of a single repeated token produces identical first-layer Value vectors across all $b$ positions of a cache block. Every row permutation therefore acts as the identity on the data matrix. Only the single-row marker shifts position. The $b!$ nominal states collapse to exactly $b$ distinguishable protected configurations, one per marker location. The attacker collects these $b$ states from its own probe blocks by observing multiple independently permuted copies of the same repeated-token input.

\begin{figure}[t]
\centering\includegraphics[width=\columnwidth]{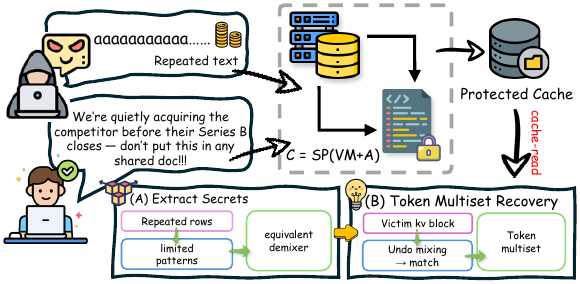}
\caption{Attack overview: repeated-input calibration followed by blockwise token recovery from protected caches.}
\label{fig:overview}
\end{figure}

We then derive an equivalent denoiser, which we call a demixer, from these $b$ configurations (Figure~\ref{fig:overview}). Because the repeated-token data is identical across all states, subtracting any two cancels it and isolates the marker's contribution, directly exposing columns of the secret row transform. The attacker then uses its own known probe vector to solve for the remaining column transform, completing the demixer. Crucially, the recovered demixer transfers to target blocks from other requests under the same secret epoch: a single calibration enables passive content recovery from any cache block protected by the same reused secrets. The attacker applies the demixer to each target block and matches the resulting rows against a public-model token dictionary, recovering a token multiset (bag of tokens) per block. A subsequent ordering step infers within-block token positions via count-constrained beam search, producing a readable text reconstruction.

We evaluate the attack on the official \scheme{} implementation across eleven public models (0.5B--32B parameters), thirteen configurations, and 3{,}000 WildChat prompts~\cite{wildchat}. 
Token-multiset recovery (bag recall $R_{\mathrm{bag}}$) is \textbf{100\%} on GPU across all configurations, including caches from a live vLLM engine. A language-model ordering step reaches mean embedding cosine $S_{\mathrm{cos}}=0.947$ on a 602-prompt content-balanced subset drawn from the same pool, recovering substantial semantic content. To quantify a concrete privacy consequence, we test sensitive-value identification on 100 synthetic cases covering six value types (passwords, card numbers, SSNs, etc.), each with 100 candidates. The program identifies the correct value with \textbf{97.5\%} expected top-1 accuracy, relying solely on the recovered token multiset without any ordering. The same orbit-collapse mechanism appears in two other shuffled-obfuscation defenses~\cite{hidden,permllm}, indicating a paradigm-level vulnerability. As a defense, we propose \veilkv{}, a byte-level authenticated-encryption scheme that binds public prefixes to scoped private tails and validates both correctness and integrity in native and vLLM execution. The developers have confirmed the vulnerability described in our detailed report submitted through their private reporting channel (Appendix~\ref{app:ethics}).

Our contributions are:
\begin{itemize}[\setlength{\IEEElabelindent}{0pt}\setlength{\labelsep}{0.6em}]
\item \textbf{Input-symmetry analysis.} We show that chosen-input symmetry reduces the observable orbit to $b$ states despite $b!$ possible permutations, exposing the reused secret parameters. The same mechanism applies to two additional public shuffled-obfuscation defenses~\cite{hidden,permllm}.
\item \textbf{Repeat-to-Reveal attack.} We derive an algebraic demixer from repeated-token probes that achieves 100\% token-multiset recovery across eleven models, thirteen configurations, optional scaling, weight fusion, and live vLLM inference.
\item \textbf{End-to-end privacy impact.} Language-model ordering recovers text at $S_{\mathrm{cos}}=0.947$ semantic similarity, and a closed-set identification program pinpoints sensitive values at 97.5\% top-1 accuracy from the recovered tokens alone.
\item \textbf{Authenticated defense.} \veilkv{} replaces obfuscation with byte-level authenticated encryption, supporting public-prefix sharing and scoped private tails with verified correctness in a running vLLM engine.
\end{itemize}

\section{Background and Threat Model}
\label{sec:model}
This section introduces the KV cache representation that enables dictionary-based content recovery (Section~\ref{sec:prelim}), formalizes the shuffled obfuscation paradigm and the specific construction we evaluate (Section~\ref{sec:construction}), and defines the attacker's capabilities and goals (Section~\ref{sec:threat}).

\subsection{KV Cache Representation}
\label{sec:prelim}
Each first-layer Value vector is a deterministic function of the input token and the public model weights, so an attacker with the same public model can precompute a token-to-vector dictionary for content recovery. Fix one first-layer Value head and one secret epoch, suppressing those indices below. For token $t$ in the public vocabulary $\mathcal V$, write
\begin{equation}
 v(t)=W_V\,N(E[t])+\beta_V,
 \label{eq:dictionary}
\end{equation}
where $E$ is the public embedding table, $N$ is the first layer's input normalization, and $W_V,\beta_V$ are the head's projection and bias. $v(t)\in\R^d$ is a column vector. The implementation detects the appropriate normalization/projection path for each model. Equation~\eqref{eq:dictionary} does not assert that deeper-layer states depend only on the token. It also does not apply to every possible architecture with absolute position-dependent inputs.

Let $b$ denote the block size and $d$ the Value-head dimension. We identify the symmetric group $\mathcal S_b$ with $b\times b$ permutation matrices and write $O(n)$ for the orthogonal group.

\subsection{Shuffled Obfuscation and the Tested Construction}
\label{sec:construction}
The shuffled obfuscation paradigm groups inference-protection designs built from secret linear transforms and permutations. A useful organizing abstraction is $C=f(P,\,\mathcal T(X)+A)$, where $\mathcal T$ is a secret invertible linear map, $P$ is a permutation, and $A$ denotes additive recovery information when present. Examples include \scheme{}~\cite{kvcloak} and permutation-based private inference~\cite{permllm,hidden}. Related constructions use refreshed batch mixing~\cite{gelo} or orthogonal conjugation~\cite{conjformer}. PermLLM and KV-Cloak cite the size of the permutation group as a barrier to brute-force recovery~\cite{permllm,kvcloak}: for $b$ shuffled elements, $P$ belongs to the symmetric group $\mathcal S_b$ of order $b!$.

We demonstrate the attack on \scheme{}, a representative instance of this paradigm. For block tokens $t_1,\ldots,t_b$, let $V\in\R^{b\times d}$ have rows $v(t_i)^\T$. The official default obfuscation path can be written as
\begin{equation}
 C=S P(VM+A),\qquad A=e_r a^\T.
 \label{eq:cloak}
\end{equation}
Here $S\in O(b)$ and $M\in O(d)$ are secret row and column transforms, and $P$ is a block-specific permutation matrix. The marker uses $r\in\{1,\ldots,b\}$, the standard basis vector $e_r\in\R^b$, and $a\in\R_{>0}^d$ in the evaluated configuration. $C,A\in\R^{b\times d}$. For a fixed head and branch, $S$, $M$, and $a$ are reused by a cloak instance. The marker's effective location changes under $P$. The implementation also exposes optional positive scaling factors. Our exact derivation first addresses the default orthogonal configuration, and we evaluate scaling settings separately.

Equation~\eqref{eq:cloak} follows the implementation's operation order: $A$ is added after the column transform and is expressed in transformed-Value coordinates. The operation order and refresh scope determine the \emph{observable orbit}, the set of distinct protected outputs that varying the block-specific $P$ alone can produce for a fixed input. When many permutations act identically on the data matrix, for instance because repeated input tokens make the Value rows equal, this orbit can be far smaller than $b!$, and the structured differences among orbit elements can expose the reused transforms $S$, $M$, and $a$ (Section~\ref{sec:analysis}).

The tested default generates $S$ by QR decomposition of a Gaussian matrix and scales a positive marker vector by the calibrated threshold. With ratio scaling disabled, $d$ is even and $M$ rotates the known coordinate pairs $I_i=(i,i+d/2)$, $i=1,\ldots,d/2$, in both the Key and Value branches. We parameterize each pair by
\begin{equation}
 M_{I_i,I_i}=R(\theta_i),\qquad
 R(\theta)=\begin{pmatrix}\cos\theta&-\sin\theta\\\sin\theta&\cos\theta\end{pmatrix},
 \label{eq:pair-rotation}
\end{equation}
with zero entries between distinct pairs (implementation details in Appendix~\ref{app:implementation-pointers}).

RoPE motivates a compatible column transform on the Key branch~\cite{kvcloak,rope}. On the Value branch, paired rotations are a property of the implementation: RoPE does not act on Value vectors. Our identification result relies on this restricted structure, because recovering a general orthogonal matrix in $O(d)$ from one known vector pair would be underdetermined.

\subsection{Threat Model}
\label{sec:threat}
\noindent Table~\ref{tab:boundary} summarizes the information available to the attacker and the information withheld by the defender. We detail each side below.

\noindent\textbf{Attacker capabilities.}
The adversary knows the public model weights and tokenizer. It can submit its own text requests and read the corresponding protected cache blocks. It can also observe a target request's protected blocks, provided that both the probe and the target are processed under the same secret configuration (the same $S$, $M$, and $a$).

A concrete instantiation is an infrastructure operator with read access to an offloaded cache pool. If the operator's own probe and a victim's request pass through the same cloak instance, the operator observes both protected caches within the required secret scope. Different cloak instances can separate requests even when they share the same public model. We assume the probe and target share a secret scope but do not measure how often real deployments satisfy this condition. Cache-level access is a prerequisite of the attack, not a contribution of it.

The adversary does not receive $S$, $M$, $a$, the victim's plaintext, or the defender's fused model weights. A public local copy of the same model supplies the token-to-vector dictionary and the probe vectors.

\noindent\textbf{Secret reuse scope.}
The tested obfuscator constructs and caches $S$, $M$, and the marker $a$ on its first invocation, then reuses these tensors for all subsequent blocks, sampling only a fresh row permutation $P$ per block. This reuse is what enables cross-request transfer: a one-time calibration on the attacker's probe yields a demixer that applies to any later target block processed under the same secrets. We assume public block size $b$ and head layout.

\noindent\textbf{Goals and exclusions.}
The primary goal is recovery of the token multiset in each complete target block. A secondary goal is closed-set identification: given an explicitly supplied candidate set of same-type values, determine which candidate appears in the block. This second goal is conditional on candidate knowledge and does not constitute unconstrained credential recovery.
The algebraic output does not assign within-block token positions. A separate language-prior reordering step (Section~\ref{sec:order-method}) addresses sequencing. We do not claim general exact-prompt reconstruction, deeper-layer recovery, or attacks on untrusted accelerators with unrestricted access to all intermediate computation.

\begin{figure*}[t]
\centering\includegraphics[width=0.8\textwidth]{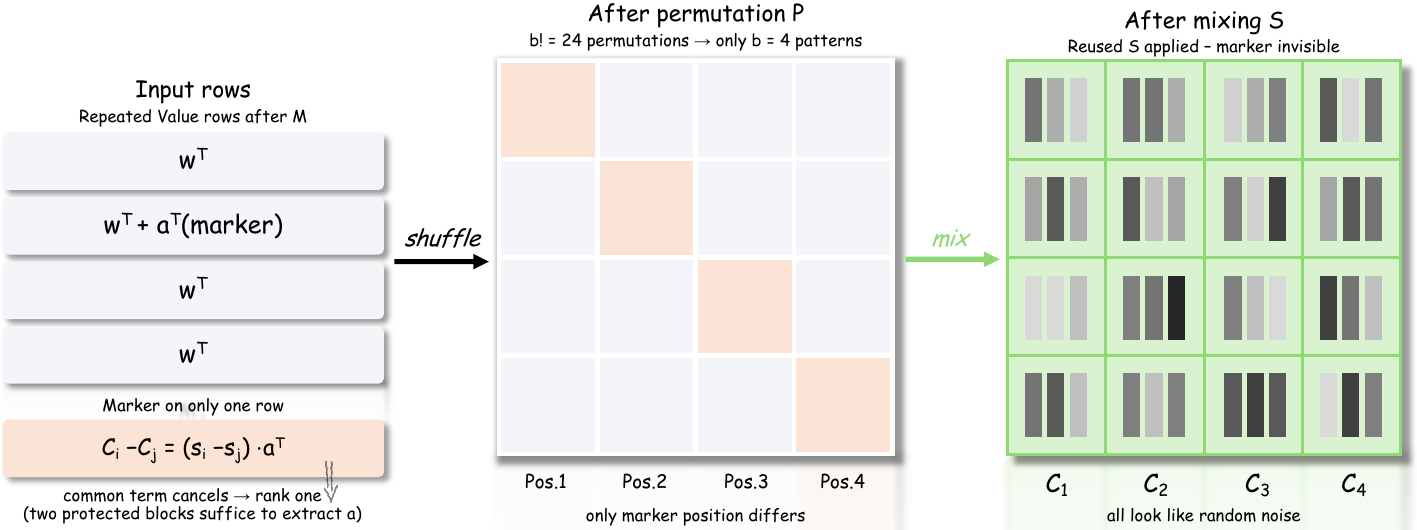}
\caption{Input symmetry with four rows. Repeated Value rows remain unchanged by permutation. Only the marker position varies. The four illustrated states are then mixed by the reused row transform.}
\label{fig:input-symmetry}
\end{figure*}

\begin{table}[t]
\centering
\caption{Information boundary used by the attack.}
\label{tab:boundary}
\begingroup
\setlength{\tabcolsep}{5pt}
\renewcommand{\arraystretch}{1}
\setlength{\extrarowheight}{1.5pt}
\begin{tabular}{>{\raggedright\arraybackslash}m{\dimexpr.5\columnwidth-2\tabcolsep-.5\arrayrulewidth\relax}|>{\raggedright\arraybackslash}m{\dimexpr.5\columnwidth-2\tabcolsep-.5\arrayrulewidth\relax}}
\TableTop
\rowcolor{TableHead}
\TableHeaderStrut \textbf{Available to attacker} & \TableHeaderStrut \textbf{Unavailable to attacker} \\
\TableHeadRule
Original model and tokenizer & Secret configuration \\
\hline
Own text and protected probe & Victim plaintext and token labels \\
\hline
Protected target blocks & Defender's fused dictionary \\
\hline
Public block/head dimensions & True per-block permutation \\
\hline
Supplied candidates (value task) & True candidate index for scoring \\
\TableBottom
\end{tabular}
\endgroup
\end{table}

\section{Security Analysis of Shuffled Obfuscation}
\label{sec:analysis}
The security of shuffled obfuscation rests on a permutation group of order $b!$, but the size of the group is not the same as the security it provides. This section asks three progressively sharper questions: how many distinguishable states can a chosen input actually produce, whether those states suffice to identify the reusable secret parameters, and what the practical collection and precision requirements are.

\noindent\textbf{Why fresh shuffling does not establish confidentiality.}
The core issue is that different permutations can have the same observable effect on a given input. When every row of the data matrix is identical, every row permutation leaves it unchanged, and the nominal $b!$-fold randomization collapses. Meanwhile, the reused transforms $S$ and $M$ preserve algebraic relationships across requests, so what the permutation fails to hide, cross-request observation can extract. This distinction applies to the entire shuffled obfuscation paradigm, not only to \scheme{}: uniform, unpredictable sampling does not remove an invariant of the sampled operation. Any defense in the paradigm therefore needs a security argument about these surviving invariants, not merely about the group order. The analysis below establishes the orbit collapse for the general paradigm, then derives the specific identification and recovery results for \scheme{}'s tested default configuration.

\subsection{Input Symmetry Collapses the Orbit}
The attacker's leverage comes from choosing the input to be obfuscated. By submitting a prompt whose tokens repeat within every block, the attacker forces all Value rows to be identical, making row permutations act as the identity on the data matrix (Figure~\ref{fig:input-symmetry}).
Formally, let $\ones_b\in\R^b$ be the all-ones vector. A repeated-token input produces $V=\ones_b v^\T$, where $v=v(t_0)$ for the probe token $t_0$, so every row permutation satisfies $PV=V$. Define $w=M^\T v\in\R^d$ and $u=S\ones_b\in\R^b$. Equation~\eqref{eq:cloak} becomes
\begin{equation}
 C(P)=u w^\T+S P A.
 \label{eq:orbit}
\end{equation}
Only the marker moves. Identical rows, rather than rank one alone, are essential: a general rank-one matrix $xv^\T$ need not be invariant when the entries of $x$ differ.

\begin{proposition}[Marker orbit]
For fixed $V=\ones_b v^\T$, fixed $A,M$, and fixed invertible $S$, the number of distinct protected outputs in exact arithmetic in~\eqref{eq:cloak} under all row permutations is
\begin{equation}
 |\{C(P):P\in\mathcal S_b\}|=\frac{b!}{|\Stab(A)|},
 \label{eq:stabilizer}
\end{equation}
where $\Stab(A)=\{P\in\mathcal S_b:PA=A\}$. For $A=e_ra^\T$ with $a\ne0$, this number is $b$.
\end{proposition}
\begin{proof}
Adding the constant $uw^\T$ and multiplying by invertible $S$ preserve equality of outputs. Thus two outputs agree exactly when their permuted markers agree. The orbit--stabilizer identity gives~\eqref{eq:stabilizer}. A single nonzero marker row can occupy $b$ distinct positions.
\end{proof}

For $k$ identical nonzero marker rows and $b-k$ zero rows, the count is $\binom bk$. If all marker rows are identical, it is one even though the marker has full support. If the $k$ nonzero rows are distinct and different from zero, the count is $b!/(b-k)!$. The orbit size therefore depends on the marker's row structure, not merely its support size. Our recovery routine targets the single-row case.

In the single-row case, let $s_j=Se_j$. The $b$ ideal states are
\begin{equation}
 C_j=uw^\T+s_j a^\T,\qquad j=1,\ldots,b.
 \label{eq:state}
\end{equation}
Each state corresponds to a different marker position. The fresh permutation $P$ is still sampled per block, but many distinct permutations map to the same state, so the attacker observes at most $b$ distinguishable outcomes instead of $b!$.

\subsection{Identification from the State Geometry}
Once the $b$ states are in hand, their geometry reveals the secret parameters. State differences $C_i - C_j$ are rank-one matrices whose singular vectors expose the marker $a$ and the columns of $S$, while the known probe vector $v$ anchors the recovery of $M$. Write $Q_a=I_d-aa^\T/\|a\|^2$ for the projection onto the orthogonal complement of $a$.

\begin{theorem}[Default-configuration identification]
\label{thm:identification}
Assume $b\ge2$, all $b$ states in~\eqref{eq:state} are available, $S\in O(b)$, $M\in O(d)$, $a\in\R_{>0}^d$ with positivity known to the attacker, and probe vector~$v$ is known.
\begin{enumerate}
\item[\textup{(i)}] If $Q_aw\ne0$ and $\langle w,a\rangle\ne-\|a\|^2/b$, then $a$, $w$, $u$ and the columns of $S$ are identifiable, with the columns unordered.
\item[\textup{(ii)}] If additionally $M$ has the paired-rotation structure in~\eqref{eq:pair-rotation} and $v_{I_i}\ne0$ for every pair, then $M$ is uniquely identifiable and each $\theta_i$ is identifiable modulo~$2\pi$.
\end{enumerate}
\end{theorem}
\begin{proof}
For $i\ne j$,
\begin{equation}
 C_i-C_j=(s_i-s_j)a^\T.
 \label{eq:rankone}
\end{equation}
This matrix has rank one. Its right singular direction identifies $a/\|a\|$ up to sign. Positivity fixes the orientation. Since distinct orthonormal columns satisfy $\|s_i-s_j\|=\sqrt2$, its nonzero singular value is $\sqrt2\|a\|$, identifying the magnitude.

The projection $C_jQ_a=u(Q_aw)^\T$ is a nonzero rank-one matrix. Its left direction identifies $u$ up to sign, and $\|u\|=\sqrt b$ fixes its magnitude. For either sign candidate, use $u^\T s_j=1$ to compute
\begin{equation}
 w^\T=\frac{u^\T C_j-a^\T}{b}.
 \label{eq:w}
\end{equation}
Since $M$ is orthogonal, $\|w\|=\|v\|$ is known. The assumed inequality makes this norm distinguish the two candidates, selecting the correct sign. Each column follows from
\begin{equation}
 s_j=\frac{(C_j-uw^\T)a}{\|a\|^2}.
 \label{eq:s}
\end{equation}
The state labels leave the original column indices unordered. For $x=v_{I_i}$ and $y=w_{I_i}$, we have $y=R(-\theta_i)x$. Each nonzero pair determines $M$'s rotation through
\begin{equation}
 \theta_i=\operatorname{atan2}(x_2y_1-x_1y_2,\,x_1y_1+x_2y_2)
 \pmod{2\pi}.\qedhere
 \label{eq:rotation-identification}
\end{equation}
\end{proof}

Two degeneracies deserve attention. Choosing $-u$ in~\eqref{eq:w} gives $w'=-w-2a/b$, not merely a harmless sign flip. The norm test resolves this ambiguity except when $\langle w,a\rangle=-\|a\|^2/b$. Similarly, a zero pair in $v$ leaves that pair's rotation undetermined. A different probe token may resolve both cases (Appendix~\ref{app:saved-record-review}).

The identification thus weaves together several structural properties: orthonormal columns supply differences of norm $\sqrt2$ that expose the marker, its known positivity fixes orientations, the public probe anchors the column rotation, and the paired-rotation structure of $M$ makes each angle separately recoverable. The resulting demixer determines the columns of $S$ up to ordering, which suffices for multiset recovery but not for within-block position claims.

\subsection{Collection Cost and Finite Precision}
The identification theorem assumes that all $b$ states have been observed. In practice, collecting them is a coupon-collector process: if each block independently maps the marker uniformly to one of $b$ rows, the expected number of blocks needed is $bH_b$, with $H_b=\sum_{i=1}^b1/i$. For $b=16$ this is about 54.1 blocks. A union bound gives
\begin{equation}
 \Pr[\text{some state missing after }n]\le b(1-1/b)^n
 \le be^{-n/b}.
 \label{eq:collection}
\end{equation}
For $H$ heads and $0<\delta<1$, $n\ge b\log(bH/\delta)$ bounds the probability of incomplete coverage by $\delta$, using a union bound over heads. In an $L$-token request, a repeat suffix starting at zero-based offset $p$ contributes $\max\{0,\lfloor L/b\rfloor-\lceil p/b\rceil\}$ usable blocks. Thus a sufficient submitted-token budget is
\begin{equation}
 L\ge b\left(\left\lceil p/b\right\rceil+
              \left\lceil b\log(bH/\delta)\right\rceil\right).
\end{equation}
The attacker can pool complete repeated-token blocks from several requests processed under the same $S,M,a$ configuration, using the same repeated token in each request. The usable block counts then add across requests, with prefix and incomplete-block loss accounted for separately in each request.

A second practical concern is floating-point noise: different accumulation orders can produce slightly different bit patterns for permutations that belong to the same ideal state. If every observed block is within Frobenius distance $\epsilon$ of its ideal state, within-state distances are at most $2\epsilon$. For distinct states, orthogonality gives
\begin{equation}
 \|C_i-C_j\|_F=\sqrt2\|a\|.
\end{equation}
Consequently a separating distance threshold exists whenever $4\epsilon<\sqrt2\|a\|$. This sufficient condition explains why rounding variants can remain clusterable. The experiments confirm recovery under the tested 32-bit floating-point (FP32) and bfloat16 (bf16) cloak operations

A similar margin argument applies to dictionary decoding. If the true normalized vector $z$ has inner-product margin $\gamma$ over every incorrect dictionary entry and the recovered vector differs from $z$ by at most $\eta$, then $2\eta<\gamma$ suffices to preserve the correct nearest neighbor. This condition governs row-level matching. Marker-hypothesis selection and vocabulary-level ambiguity are measured empirically.

\section{Attack Construction}
\label{sec:attack}

Section~\ref{sec:analysis} showed how repeated inputs expose $b$ protected states and how those states identify the reused secret parameters. Algorithm~\ref{alg:attack} turns this analysis into a recovery procedure. Phase~1 uses the attacker's repeated-token probe to estimate the transforms for each Value head. This calibration is performed once and reused for other requests protected with the same secrets. Phase~2 recovers which tokens occur in each target block and how often they occur. Phase~3 uses a public language model to infer their order within each block. The first output is a block token multiset, and the second is an inferred token order.

\begin{algorithm}[!t]
\caption{Repeated-token attack on shuffled cache obfuscation}
\label{alg:attack}
\begingroup\small
\begin{algorithmic}[1]
\Statex \textbf{Input:} public model weights, tokenizer, block size~$b$, head dim~$d$
\Statex \textbf{Ordering settings:} beam width $\beta$, context limit $c$
\Statex \textbf{Output:} block token multisets and inferred prompt text
\Statex
\Statex \textit{// Phase 1: Calibration (per head)}
\State Find probe token $t_0$ yielding a long constant-token run
\State Submit probe text and collect protected blocks $C^{(1)},\ldots,C^{(n)}$
\For{each Value head}
  \State Compute pairwise distances and set threshold~$\tau$
  \State Cluster blocks into $b$ centroids $\widehat C_1,\ldots,\widehat C_b$
  \State \textbf{if} fewer than $b$ clusters \textbf{then} skip head
  \State Extract marker $\widehat a$ via SVD of $\widehat C_i-\widehat C_j$ \hfill\Comment{Eq.~\eqref{eq:rankone}}
  \State Fix sign: require $\sum_\ell \widehat a_\ell>0$
  \State Project $\widehat C_j Q_{\widehat a}$ to recover $\widehat u$ \hfill\Comment{left singular dir.}
  \State Recover $\widehat w$ from $\widehat u,\widehat C_j,\widehat a$ \hfill\Comment{Eq.~\eqref{eq:w}}
  \State Recover columns $\widehat s_j$ for $j=1,\ldots,b$ \hfill\Comment{Eq.~\eqref{eq:s}}
  \State Recover rotation angles $\widehat\theta_i$ from $(v_{I_i},\widehat w_{I_i})$ \hfill\Comment{Eq.~\eqref{eq:rotation-identification}}
\EndFor
\State Assemble $\widehat S$, $\widehat M$ from recovered parameters
\Statex
\Statex \textit{// Phase 2: Recover multiset}
\State Build public dictionary $\{v(t):t\in\mathcal V\}$ \hfill\Comment{Eq.~\eqref{eq:dictionary}}
\For{each protected target block $C$}
  \For{$j=1,\ldots,b$}
    \State $\widehat Y_j\gets\widehat S^\T(C-\widehat s_j\widehat a^\T)\widehat M^\T$ \hfill\Comment{Eq.~\eqref{eq:decode}}
    \State Match each row of $\widehat Y_j$ to nearest dictionary entry
    \State $\mathrm{score}_j\gets\sum_i\max_t\cos(\widehat Y_{j,i},\,v(t))$
  \EndFor
  \State Select $j^*=\argmax_j\mathrm{score}_j$
  \State Store the token multiset from $\widehat Y_{j^*}$
\EndFor
\Statex
\Statex \textit{// Phase 3: Infer order}
\For{each target request}
  \State Initialize empty text $\widehat y$ and start context $h$
  \For{each block multiset $\mathcal B$ in request order}
    \State Init beam: empty sequence, token counts from $\mathcal B$
    \For{$\ell=1,\ldots,b$}
      \State Extend candidates with tokens still available in $\mathcal B$
      \State Update counts and scores \hfill\Comment{Eq.~\eqref{eq:order-score}}
      \State Retain up to $\beta$ highest-scoring candidates
    \EndFor
    \State Append best block to $\widehat y$ and update $h$ to last $c$ tokens
  \EndFor
  \State Output the block multisets and inferred text $\widehat y$
\EndFor
\end{algorithmic}
\endgroup
\end{algorithm}

\subsection{Phase 1: Probe Construction and Calibration}
\label{sec:calibration}

\noindent\textbf{Probe construction (lines 1--2).}
The attacker constructs a prompt that produces a long constant-token run after tokenization. If this run occupies the suffix $[p,L)$ of the $L$-token input (zero-based), the usable block indices are
\begin{equation}
 \left\lceil p/b\right\rceil,\ldots,\left\lfloor L/b\right\rfloor-1.
\end{equation}
The full tokenizer output, including any prefix tokens, is submitted. The attacker filters to the constant-suffix region after collection.

\noindent\textbf{Adaptive clustering (lines 4--6).}
Because the orbit has only $b$ states, the probe blocks should separate into $b$ tight clusters, one per marker position. The routine finds them with an adaptive distance threshold: let $d_{\min}$ and $d_{\max}$ be the smallest positive and largest off-diagonal Euclidean distances among flattened probe blocks. Then $\tau=\min\{\sqrt{d_{\min}d_{\max}},d_{\max}/2\}$, or $1$ when no positive distance exists. If fewer than $b$ clusters emerge, the head is skipped.

\noindent\textbf{Parameter recovery (lines 7--14).}
With $b$ cluster centroids in hand, the attacker applies the identification procedure from Theorem~\ref{thm:identification}. Write the marker estimate as $\widehat a=\widehat\alpha\widehat q_a$, where $\widehat q_a$ is a unit direction. For distinct ideal states, $\|a\|=\sigma_1(C_i-C_j)/\sqrt2$. The corrected calibrator normalizes the averaged marker direction before forming the projection and aligns the left singular vectors before averaging them. We denote the estimated common vectors by $\widehat u,\widehat w$, the recovered columns by $\widehat s_j$, and the corresponding matrices by $\widehat S=[\widehat s_1,\ldots,\widehat s_b]$ and $\widehat M$. Under exact recovery, $\widehat S=S\Pi$ for an unknown permutation $\Pi$. This column reordering does not affect content extraction, since it only changes the order in which recovered rows are returned.

\subsection{Phase 2: Token-Multiset Recovery}
\label{sec:decoding}

\noindent\textbf{Dictionary construction (line 15).}
The attacker precomputes a dictionary mapping each vocabulary token to its first-layer Value vector via Equation~\eqref{eq:dictionary}, using only the public model weights. The implementation validates dictionary consistency against locally computed probe states.

\noindent\textbf{Demixing and hypothesis selection (lines 16--23).}
For marker hypothesis $j$, the demixer produces
\begin{equation}
 \widehat Y_j=\widehat S^\T(C-\widehat s_j\widehat a^\T)\widehat M^\T.
 \label{eq:decode}
\end{equation}
Under the correct hypothesis and exact recovery, $\widehat Y_j=\Pi^\T PV$ is a row-permuted copy of~$V$, so each row should match a dictionary entry closely. The hypothesis with the largest total cosine score is selected. Success depends on the vocabulary's discriminability for each block (Appendix~\ref{app:saved-record-review}).

\noindent\textbf{Multiset scoring.}
The algebraic decoding step recovers token identities and repetition counts, leaving their within-block order unknown. Let $\mathcal B$ and $\widehat{\mathcal B}$ be the true and recovered token multisets, and let $n_{\mathcal B}(t)$ count occurrences of token $t$. We score their multiset intersection:
\begin{equation}
 \operatorname{hit}(\widehat{\mathcal B},\mathcal B)
 =\sum_{t\in\mathcal V}\min\{n_{\widehat{\mathcal B}}(t),n_{\mathcal B}(t)\}.
 \label{eq:metric}
\end{equation}
The denominator is $b$ per scored head and block. This token-recovery step tries $b$ marker hypotheses rather than $b!$ permutations. Its dictionary-matching cost therefore depends on vocabulary size and nearest-neighbor search (Appendix~\ref{app:cpu-costs}).

\subsection{Phase 3: Token Order Reconstruction}
\label{sec:order-method}
The recovered multisets fix the token identities and repetition counts within each block, while the request's block order remains known. The attacker infers within-block order using a public autoregressive language model over the same token vocabulary~\cite{wordorder,ibis}. Let $h$ denote the previously reconstructed context and $z=(z_1,\ldots,z_b)$ a candidate sequence whose token multiset equals the recovered $\mathcal B$. The model's next-token probabilities $p_{\mathrm{LM}}$ give the score
\begin{equation}
 \operatorname{score}_{\mathrm{LM}}(z\mid h)
 =\sum_{\ell=1}^{b}\log p_{\mathrm{LM}}(z_\ell\mid h,z_{<\ell}).
 \label{eq:order-score}
\end{equation}
We approximate the highest-scoring admissible order with count-constrained beam search of width $\beta$. Each candidate maintains a partial sequence, its cumulative score and the remaining token counts. At each step, the search extends each candidate with every token whose remaining count is positive, decrements that count and updates the score. It retains at most $\beta$ candidates with the highest scores. After $b$ steps, the best remaining sequence is appended to the reconstructed prompt. The context for the next block contains up to $c$ most recent tokens of the reconstructed prompt, where $c$ is the context-length limit. The first block uses a public start token. This preserves each block's recovered multiset while inferring its internal order.

\section{Evaluation}
\label{sec:evaluation}
We evaluate the orbit collapse and the Repeat-to-Reveal attack on \scheme{} and confirm the shared row-symmetry mechanism on two additional public implementations~\cite{hidden,permllm}. Across eleven models (0.5B--32B), thirteen configurations and 3{,}000 WildChat prompts, the attack achieves 100\% token-multiset recovery on GPU, including caches produced during live vLLM inference. Beam-search reordering of the recovered tokens reaches a mean sentence-embedding cosine of 0.947 with the originals, and sensitive-value identification achieves 97.5\% expected top-1 accuracy. The evaluation progresses from mechanism validation through attack effectiveness and privacy impact to defense, guided by four research questions:

\textbf{RQ1} (Mechanism validation): Does the orbit collapse predicted manifest empirically, and do the resulting states enable accurate calibration? \textbf{RQ2} (Attack effectiveness): How completely can the recovered demixer reconstruct target content and order, and when does calibration transfer across requests? \textbf{RQ3} (Sensitive-information leakage): Can recovered tokens reveal specific sensitive values in a user's request? \textbf{RQ4} (Defense practicality): Can \veilkv{} preserve correct inference and authorized cache reuse at acceptable cost?

\subsection{Experimental Methodology}
\label{sec:methodology}
\noindent\textbf{Implementation and models.}
We evaluate the official \scheme{} protection implementation (commit \textnormal{6b40f36}) using a single NVIDIA A100-SXM4-80GB GPU, PyTorch 2.6.0+cu124, and Transformers 4.51.3. The main suite uses 32-bit floating-point (FP32) precision for model inference and cache protection. Calibration runs on the CPU, while target decoding runs on the GPU. To test recovery across architectures and block sizes, we use eleven public models from the Qwen, Llama, OLMo and TinyLlama families~\cite{qwen25,qwen3,llama3,olmo2,tinyllama}. All models use 16-token blocks, and we additionally evaluate Qwen2.5-0.5B with 8- and 32-token blocks, resulting in thirteen configurations in total. We set \scheme{}'s marker scale from the K/V magnitudes observed on ordinary public text.

\noindent\textbf{Target prompts.}
The recovery targets are user prompts, distinct from the repeated-token calibration probes. From \textnormal{allenai/WildChat-4.8M}~\cite{wildchat,wildchat48m}, we take the first user message in each conversation, keep English messages from 2025 conversations, and exclude messages marked as toxic or redacted. To cover a range of request lengths, we measure sampling lengths with the Qwen2.5-0.5B tokenizer and randomly select 600 prompts from each of five token-length groups (32--63, 64--127, 128--255, 256--511, 512--767), giving 3{,}000 prompts, and duplicates are excluded. Every randomized step uses a fixed seed. Each model tokenizes these texts with its own tokenizer and prefixes. Because the ordering and ablation experiments require per-prompt language-model decoding, we use a fixed subset of 602 prompts selected by content category (Appendix~\ref{app:subset602}). All ordering, secret-rotation, scaling, weight-fusion, and live-vLLM experiments use this same fixed subset. The primary recovery result ($R_{\mathrm{bag}}=100\%$) is evaluated on the full 3{,}000 prompts. The sensitive-value study uses separately constructed synthetic cases.

\noindent\textbf{Attack procedure.}
For each model and block size, we calibrate the attack once and apply the resulting estimates to targets protected under the same secret configuration. The true secret parameters are used only to evaluate estimation accuracy. Each target is protected twice with independently sampled block permutations while all other parameters remain fixed, isolating the effect of fresh permutations. No prior attack recovers tokens from dense-mixed shuffled caches: Thomas et al.~\cite{hidden} target permutation-only defenses, making a direct comparison inapplicable(Appendix~\ref{sec:arithmetic-profile}). 

\noindent\textbf{Metrics.}
We separately measure parameter accuracy (RQ1), token-multiset recovery and ordering (RQ2), sensitive-value identification (RQ3), and cache correctness with runtime overhead (RQ4). Table~\ref{tab:metrics} defines all metrics. Rates are percentages unless shown as exact/total counts.

\begin{table*}[!t]
\centering
\caption{Evaluation metrics.
$\operatorname{hit}(\widehat{\mathcal B},\mathcal B)$ counts matching token occurrences between two multisets (Eq.~\eqref{eq:metric}). $\mathcal I$ indexes all blocks across targets, configurations, KV heads and shuffle seeds. Here $Q$ is the number of evaluated targets. For sequence metrics, $y_j$ and $\widehat y_j$ denote the reference and reconstructed token sequences for prompt $j$, and $L_j$ is the reference length.}
\label{tab:metrics}
\begingroup\footnotesize
\setlength{\tabcolsep}{4pt}
\renewcommand{\arraystretch}{1.05}
\setlength{\extrarowheight}{0pt}
\resizebox{0.98\textwidth}{!}{%
\begin{tabular}{@{} m{1.7cm} | p{1.8cm} | p{13cm} @{}}
\TableTop
\rowcolor{TableHead}
\TableHeaderStrut \textbf{Aspect} & \textbf{Metric} & \textbf{Definition} \\
\TableHeadRule
Parameter (RQ1)
  & min column cosine & Minimum cosine similarity over matched recovered and true columns of $S$, where $1{=}$perfect direction match. \\
\cline{2-3}
  & rel.\ error & $\|\widehat w{-}w\|_2/\|w\|_2$.\enspace Relative estimation error of the recovered vector ($0{=}$perfect). \\
\cline{2-3}
  & sep.\ ratio & Smallest above-threshold\,/\,largest below-threshold Frobenius distance between probe blocks. Larger values indicate clearer separation. \\
\hline
Token content (RQ2)
  & bag recall & $R_{\mathrm{bag}}={\sum_{i\in\mathcal I}\operatorname{hit}(\widehat{\mathcal B}_i,\,\mathcal B_i)}\big/{\sum_{i\in\mathcal I}|\mathcal B_i|}$.\enspace Pooled bag recall~\cite{rajpurkar2016squad}, equal to micro-averaged $F_1$ when each predicted block has the same length as its reference (App.~\ref{app:metrics}).\hfill\refstepcounter{equation}\label{eq:pooled-recovery}(\theequation) \\
\cline{2-3}
  & bag $F_1$ & $F_1={2\,\operatorname{hit}(\widehat{\mathcal B},\,\mathcal B)}\big/({|\widehat{\mathcal B}|+|\mathcal B|})$.\enspace Per-bag harmonic mean of precision and recall. \\
\hline
Token order (RQ2)
  & position acc. & $A_{\mathrm{pos}}=\frac{1}{Q}\sum_{j}\frac{1}{L_j}\sum_{i}\mathbf{1}[\widehat y_{j,i}{=}y_{j,i}]$.\enspace Mean per-prompt fraction of tokens placed at the correct position~\cite{hidden,morris2023}.\hfill\refstepcounter{equation}\label{eq:sequence-metrics}(\theequation) \\
\cline{2-3}
  & exact match & $\mathrm{EM}=\frac{1}{Q}\sum_{j}\mathbf{1}[\widehat y_j{=}y_j]$.\enspace Fraction of fully correctly ordered sequences. \\
\cline{2-3}
  & embedding cos. & $S_{\mathrm{cos}}=\frac{1}{Q}\sum_{j}\cos\bigl(e(y_j),\,e(\widehat y_j)\bigr)$.\enspace Mean sentence-embedding cosine similarity, where $e$ is all-MiniLM-L6-v2~\cite{morris2023} with a limit of 256 encoder tokens.\hfill\refstepcounter{equation}\label{eq:semantic-metric}(\theequation) \\
\hline
Sensitive-value ID (RQ3)
  & candidate $F_1$ & $s_r=\frac{1}{2}\sum_{h}\frac{2|\mathcal B_r\cap\widehat{\mathcal B}_h|}{|\mathcal B_r|+|\widehat{\mathcal B}_h|}$.\enspace F1 averaged over KV heads, where $\mathcal B_r$ is the candidate multiset and $\widehat{\mathcal B}_h$ the recovery from head $h$.\hfill\refstepcounter{equation}\label{eq:candidate-score}(\theequation) \\
\cline{2-3}
  & top-$k$ acc. & $\operatorname{Acc}@k=\frac{1}{Q}\sum_{j}p_{j,k}$,\enspace $p_{j,k}{=}\min\{1,\,\max\{0,\,(k{-}g_j)/m_j\}\}$.\enspace Top-$k$ accuracy with tie-breaking, where $g_j$ counts candidates scored above truth and $m_j$ the tied count.\hfill\refstepcounter{equation}\label{eq:ties}(\theequation) \\
\hline
Defense (RQ4)
  & byte match & Recovered cache has identical data types and byte values as the unprotected reference. \\
\cline{2-3}
    & continuation & Exact logits and generated tokens for \veilkv{}. Identical predicted next tokens and logit errors within the stated tolerances for the $S,a$ study (Appendix~\ref{app:sa-continuation}). \\
\cline{2-3}
  & runtime & Median wall-clock time in milliseconds per call, reported separately for each length group. \\
\TableBottom
\end{tabular}}%
\endgroup
\end{table*}

\subsection{RQ1: Mechanism Validation}
\label{sec:rq1}
This section validates the orbit-collapse mechanism underlying the attack. The attack requires the repeated-token probe to produce distinguishable protected states whose geometry reveals the secret parameters. We confirm that the predicted $b$ states emerge with clear separation and support accurate calibration, determine the minimum probe length needed to observe all states, and verify that the same row-symmetry mechanism appears in two other shuffled-obfuscation defenses.

\noindent\textbf{From states to parameters.}
After removing prefix blocks, every evaluated head forms exactly $b$ clusters (8, 16 or 32, matching the block size), as predicted by the marker-orbit analysis. Cluster separation is unambiguous: the sep.\ ratio exceeds 1{,}324.6 in every KV head despite floating-point noise. From the cluster centroids we recover the row-mixing matrix $S$ and the transformed probe Value vector $w$ (Section~\ref{sec:calibration}). The minimum matched-column cosine for $S$ is 0.99999988, and the rel.\ error for $w$ is at most $3.79\times10^{-6}$. These results confirm accurate parameter recovery from the observed states.

\begin{figure}[t]
\centering
\includegraphics[width=0.8\columnwidth]{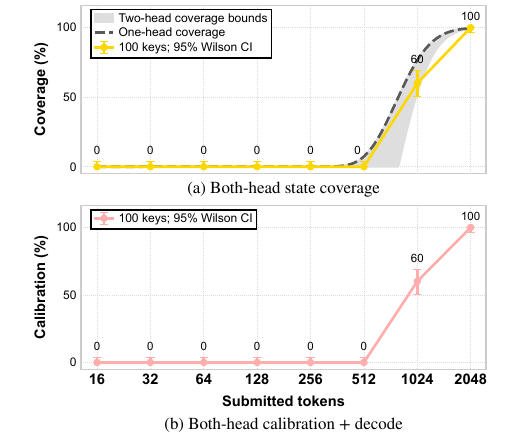}
\caption{Calibration success rate vs.\ probe length over 100 secret configurations. (a) Full state coverage: whether the probe reveals all $b$ states in both KV heads. (b) Successful calibration: whether the recovered parameters pass accuracy checks and exactly recover probe tokens. Error bars are 95\% Wilson CIs. The shaded band gives the coverage probability under independent uniform shuffles, allowing inter-head correlation (Appendix~\ref{app:bounds}).}
\label{fig:probe}
\end{figure}

\noindent\textbf{Calibration cost.}
We vary probe length across eight budgets on 100 Qwen2.5-0.5B configurations ($b=16$), reusing the same secret keys (Figure~\ref{fig:probe}). A probe of $L$ tokens without a prefix supplies $n=L/16$ blocks. We evaluate two success criteria: full state coverage, meaning the probe reveals all 16 states in both heads, and successful calibration, meaning that both the marker cosine and the minimum matched-column cosine are at least 0.999 in each head, column matching is one-to-one, and the probe tokens are recovered exactly.

At 512 tokens or fewer, neither criterion is met in any of the 100 trials. At 1{,}024 tokens, both succeed in 60/100 trials. At 2{,}048 tokens, both succeed in all 100. Every trial that achieves full coverage also calibrates successfully, indicating that the bottleneck is collecting all states, not the parameter recovery step itself.

\noindent\textbf{Row symmetry in other shuffled-obfuscation defenses.}
To test whether the orbit collapse extends beyond \scheme{}, we submit the same repeated-token probes to two other public defenses within the shuffled-obfuscation paradigm. Each probe consists of a single beginning-of-sequence (BOS) token followed by repetitions of one token, producing identical embedding rows except for the BOS position.

In Thomas et al.'s sequence-permutation defense~\cite{hidden}, permuting the repeated rows leaves them unchanged, so only the BOS row moves, yielding exactly $b$ states across four block sizes in both FP32 and bf16. A control that also shuffles vector coordinates retains distinct outputs. PermLLM~\cite{permllm} additionally shuffles coordinates within each row, keeping raw outputs distinct. However, sorting the values within each row removes coordinate-order variation and reveals only $b$ distinguishable row arrangements. The same limited orbit structure therefore appears in all three defenses. All probes passed through PermLLM's native three-party LayerNorm protocol remain stable.

\begin{answerBox}
\textbf{Answer to RQ1.} A 2{,}048-token probe reveals all states and supports accurate calibration in all tested configurations. Repeated inputs also expose limited row structure in two other shuffled-obfuscation defenses, confirming the broader input-symmetry mechanism.
\end{answerBox}

\subsection{RQ2: Attack Effectiveness}
\label{sec:rq2}
With the orbit-collapse mechanism confirmed, we now evaluate how completely the attack reconstructs target content. We proceed in two stages: first recovering the token multiset in each block (bag recall $R_{\mathrm{bag}}$), then inferring within-block token order (position acc.\ $A_{\mathrm{pos}}$, embedding cos.\ $S_{\mathrm{cos}}$). We further test dependence on secret reuse, implementation variants (weight fusion, optional scaling), and live vLLM inference.

\subsubsection{Token-Multiset Recovery and Transfer}

\noindent\textbf{Recovery on user prompts.}
Using the calibrated parameters, we recover tokens from target requests protected under the same secret configuration. Across all thirteen model configurations and 3{,}000 prompts, recovery is perfect: $R_{\mathrm{bag}}=100\%$ ($116{,}571{,}072/116{,}571{,}072$ multiset hits, Table~\ref{tab:breadth}). Two independently shuffled copies per target confirm that fresh block permutations do not prevent recovery. The full Qwen3-14B model runs in FP32, while Qwen3-32B uses only its first layer to fit within a single GPU's memory.

\begin{table*}[!t]
\centering
\caption{GPU first-layer Value recovery on the same 3,000 prompts, all KV heads and two shuffle seeds.}
\label{tab:breadth}
\setlength{\tabcolsep}{6pt}
\newlength{\BreadthDataWidth}
\setlength{\BreadthDataWidth}{\dimexpr\textwidth-14\tabcolsep-2\arrayrulewidth\relax}
\begin{tabular}{>{\centering\arraybackslash}m{0.22\BreadthDataWidth}|>{\centering\arraybackslash}m{0.06\BreadthDataWidth}>{\centering\arraybackslash}m{0.08\BreadthDataWidth}>{\centering\arraybackslash}m{0.08\BreadthDataWidth}>{\centering\arraybackslash}m{0.12\BreadthDataWidth}|>{\centering\arraybackslash}m{0.30\BreadthDataWidth}>{\centering\arraybackslash}m{0.14\BreadthDataWidth}}
\TableTop
\rowcolor{TableHead}
\TableHeaderStrut \multirow{2}{*}{\TableHeaderStrut \textbf{Model}} & \multicolumn{4}{c|}{\TableHeaderStrut \textbf{Configuration}} & \multicolumn{2}{c}{\TableHeaderStrut \textbf{Token recovery}} \\
\cline{2-7}
 & $\boldsymbol{b}$ & \TableHeaderStrut \textbf{Heads} & $\boldsymbol{d}$ & \TableHeaderStrut \textbf{Probe} & \TableHeaderStrut \textbf{Matches/total} & \TableHeaderStrut \textbf{\boldmath$R_{\mathrm{bag}}$ (\%)} \\
\TableHeadRule
Llama-3.2-1B & 16 & 8 & 64 & 3,201 & 12,307,712/12,307,712 & \RateCell{100.00}\\
\hline
OLMo-2-1B & 16 & 16 & 128 & 3,200 & 24,748,544/24,748,544 & \RateCell{100.00}\\
\hline
\multirow{3}{*}{Qwen2.5-0.5B} & 16 & 2 & 64 & 3,200 & 3,121,856/3,121,856 & \RateCell{100.00}\\
\cline{2-7}
 & 32 & 2 & 64 & 20,250 & 3,025,920/3,025,920 & \RateCell{100.00}\\
\cline{2-7}
 & 8 & 2 & 64 & 1,280 & 3,167,744/3,167,744 & \RateCell{100.00}\\
\hline
Qwen2.5-1.5B & 16 & 2 & 128 & 3,200 & 3,121,856/3,121,856 & \RateCell{100.00}\\
\hline
Qwen2.5-3B & 16 & 2 & 128 & 3,200 & 3,121,856/3,121,856 & \RateCell{100.00}\\
\hline
Qwen2.5-7B & 16 & 4 & 128 & 3,200 & 6,243,712/6,243,712 & \RateCell{100.00}\\
\hline
Qwen3-0.6B & 16 & 8 & 128 & 3,200 & 12,480,000/12,480,000 & \RateCell{100.00}\\
\hline
Qwen3-1.7B & 16 & 8 & 128 & 3,200 & 12,480,000/12,480,000 & \RateCell{100.00}\\
\hline
TinyLlama-1.1B & 16 & 4 & 64 & 2,033 & 7,791,872/7,791,872 & \RateCell{100.00}\\
\hline
Qwen3-14B & 16 & 8 & 128 & 3,200 & 12,480,000/12,480,000 & \RateCell{100.00}\\
\hline
Qwen3-32B & 16 & 8 & 128 & 3,200 & 12,480,000/12,480,000 & \RateCell{100.00}\\
\TableHeadRule
\rowcolor{TableBlue}
\multicolumn{5}{r}{\textbf{Total}} & \textbf{116,571,072/116,571,072} & \RateShade{100.000}{\textbf{100.000}} \\
\TableBottom
\end{tabular}
\par\smallskip
\parbox{\textwidth}{\textit{Note.} $b$: tokens per block, $d$: dimensions per Value head, Heads: number of KV heads, Probe: calibration-probe length in tokens after tokenization. Matches/total counts matching and evaluated token occurrences in $R_{\mathrm{bag}}$~\eqref{eq:pooled-recovery}. Total aggregates all eleven models and thirteen configurations.}
\end{table*}

\noindent\textbf{Dependence on the secret configuration.}
Recovery does not rely on a favorable key: across five independently seeded configurations, both Qwen2.5-0.5B and Llama-3.2-1B achieve 100\% bag recall on every prompt. Calibration, however, is tied to the current secrets. On 602 fixed prompts (359{,}936 token occurrences), replacing $S$, $M$ and $a$ while retaining the old estimates drops recovery to 0.412\%, but a fresh calibration restores 100\%. Fresh per-block permutations alone do not invalidate the calibration because $S$, $M$ and $a$ remain fixed (Appendix~\ref{app:matched-controls}).

\noindent\textbf{Fusing the transform into model weights.}
\scheme{} can fold the column transform $M$ into the Value-projection weights, producing $VM$ directly. The attacker, using the original public model, estimates the inverse from its own probe. On 602 Qwen2.5-0.5B targets this achieves $359{,}936/359{,}936$ recovered tokens (100\%), showing that weight fusion does not prevent recovery.

\begin{figure}[t]
\centering
\includegraphics[width=0.8\columnwidth]{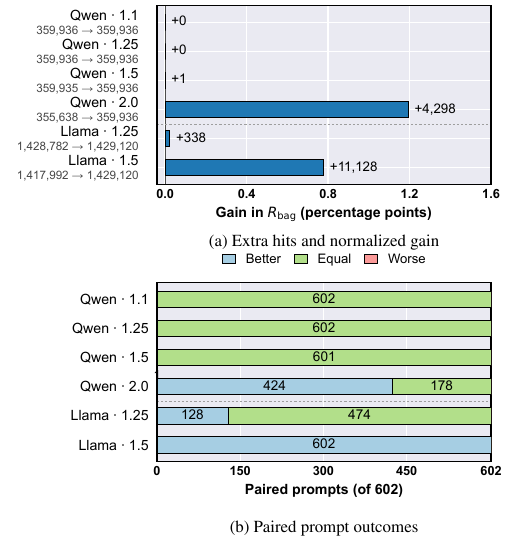}
\caption{Six scaling settings on 602 WildChat prompts. The top panel reports the change in $R_{\mathrm{bag}}$ after fitting the scaling parameter, together with the additional correctly recovered token occurrences across all evaluated heads. The bottom panel reports how many prompts improve, remain unchanged or become worse. Labels retain the correct counts before and after fitting. Denominators are 359,936 for Qwen and 1,429,120 for Llama.}
\label{fig:scaling-followup}
\end{figure}

\noindent\textbf{Additional scaling.}
\scheme{} optionally adds scaling to its default rotations. Since Theorem~\ref{thm:identification} assumes rotation-only transforms, we adapt the estimation step to account for scaling. Under all six tested settings on Qwen2.5-0.5B and Llama-3.2-1B (602 prompts each), the adapted decoder achieves 100\% bag recall, with recovery improving in four settings and unchanged in the other two (Figure~\ref{fig:scaling-followup}). Scaling therefore does not block the attack.

\noindent\textbf{Recovery from a running engine.}
We integrate the attack with vLLM 0.8.5.post1 to test recovery from caches produced during live inference. The engine processes each request, \scheme{} protects its cache, and the attack reads the protected Value vectors. Across two KV heads on 602 prompts, all token occurrences are recovered (100\%), extending the result from offline caches to a running serving engine (Appendix~\ref{app:vllm-live}).

\subsubsection{From Multisets to Ordered Text}
The recovered multisets fix token identities and counts but leave within-block order unknown. We infer order with the language-model beam search of Section~\ref{sec:order-method}.

\noindent\textbf{Recovering token positions.}\label{sec:ordering}
We order 16-token blocks recovered from head~0 of Qwen2.5-0.5B under FP32 protection, using a bf16 Qwen2.5-0.5B language-model prior, beam width 16 and up to 80 tokens of previously reconstructed context. On 602 prompts we compare four conditions: (1) LM-ordered recovered tokens with context, (2) LM-ordered true tokens (positions hidden) with context, (3) LM-ordered recovered tokens without context, and (4) random within-block shuffle. All conditions preserve each block's token counts.

Conditions (1) and (2) yield identical results ($A_{\mathrm{pos}}=28.638\%$) because multiset recovery is exact for all 602 prompts (Table~\ref{tab:ordering}). Removing context (3) or using random order (4) both lower position acc., confirming that the language model and cross-block context each contribute. Even so, substantial ordering ambiguity remains: only 1/602 sequences is fully correct (Appendix~\ref{app:ordering}).

\begin{table}[t]\centering\caption{Reconstruction on 602 prompts: position acc.\ $A_{\mathrm{pos}}$ (\%), exact match EM (exact/total)~\eqref{eq:sequence-metrics}, and embedding cos.\ $S_{\mathrm{cos}}$~\eqref{eq:semantic-metric}. Context is previously reconstructed text from preceding blocks. True tokens are supplied without their order.}\label{tab:ordering}\setlength{\tabcolsep}{3pt}
\begin{tabular}{>{\raggedright\arraybackslash}m{0.47\columnwidth}|>{\centering\arraybackslash}m{0.17\columnwidth}>{\centering\arraybackslash}m{0.13\columnwidth}>{\centering\arraybackslash}m{0.12\columnwidth}}
\TableTop
\rowcolor{TableHead}
\TableHeaderStrut \textbf{Ordering setting} & \shortstack{\TableHeaderStrut\textbf{Position}\\\textbf{\boldmath$A_{\mathrm{pos}}$ (\%)}} & \textbf{\boldmath$\mathrm{EM}$} & \shortstack{\TableHeaderStrut\textbf{Cosine}\\\textbf{\boldmath$S_{\mathrm{cos}}$}} \\
\TableHeadRule
Recovered tokens with context & \RateCell{28.638} & 1/602 & 0.94725 \\
True tokens with context & \RateCell{28.638} & 1/602 & 0.94725 \\
\hline
Recovered tokens, block alone & \RateCell{11.662} & 0/602 & 0.92719 \\
Random ordering within blocks & \RateCell{7.732} & 0/602 & 0.81713 \\
\TableBottom
\end{tabular}
\end{table}

Beam reconstruction reaches embedding cos.\ $S_{\mathrm{cos}}=0.947$ with the originals, substantially above random ordering of the same tokens ($S_{\mathrm{cos}}=0.817$, Table~\ref{tab:ordering}, Figure~\ref{fig:ordering-distributions}). The language-model ordering thus recovers much of the semantic content beyond what token identities alone reveal, suggesting that the combination of exact multiset recovery and LM-based reordering may pose a concrete privacy risk.

\begin{answerBox}
\textbf{Answer to RQ2.} With separate calibration for each model, block size and secret configuration, recovery reaches $R_{\mathrm{bag}}=100\%$, including in live vLLM inference. Each calibration transfers across requests sharing the same secrets. Beam-search ordering reaches embedding cos.\ $S_{\mathrm{cos}}=0.947$, revealing substantial prompt content despite imperfect position recovery.
\end{answerBox}

\subsection{RQ3: Sensitive-Information Leakage}
\label{sec:rq3}
RQ2 establishes that the attack recovers complete token multisets and reconstructs text with high semantic similarity. We now ask a more targeted privacy question: can the recovered tokens identify a specific sensitive value, such as a password or card number, embedded in a user's request? This closed-set identification task (Section~\ref{sec:threat}) does not require token ordering. We implement it as an automated program that ranks candidate values by their candidate $F_1$ scores~\eqref{eq:candidate-score}. We evaluate identification using expected top-$k$ acc.~\eqref{eq:ties}.

\noindent\textbf{Task setup.}
The attacker knows the surrounding request text and is given a list of 100 candidate values of the same type. One candidate is the true value embedded in the request. This closed-set setting follows prior cache-leakage studies that examine private inputs within known request formats~\cite{promptpeek2025,earlybird}.

We construct 100 synthetic test cases covering six sensitive-value types: passwords, credit-card numbers, Social Security numbers, medical record numbers, personal names, and postal addresses. Each case embeds one true value in a realistic request template (e.g., \emph{``My password is [value]. Please update it.''}), with 100 distinct candidates from a fixed pool for that value type. The surrounding text stays fixed across candidates so that only the private value differs. Synthetic requests provide a known ground truth and isolate the private value from lexical variation. The candidate lists are presented in random order. All cases use Qwen2.5-0.5B.

\noindent\textbf{Scoring procedure.}
For each candidate, the program inserts it into the template, tokenizes the full request, and compares the resulting tokens with those recovered from the cache, using only complete cache blocks. Because the scored cache blocks contain both the private value and known surrounding words, candidates whose private-value tokens better match the recovered multiset receive higher scores. The program computes candidate $F_1$~\eqref{eq:candidate-score} against each of the two KV heads separately, averages the two scores, and selects the highest-scoring candidate. No reconstruction of token order is needed for this ranking.

\begin{figure}[t]
\centering
\includegraphics[width=0.8\columnwidth]{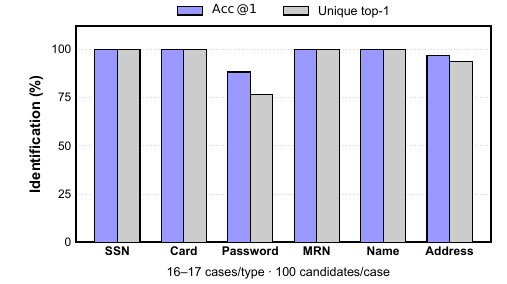}
\caption{Sensitive-value identification from 100 candidates per test. Top-$1$ acc.\ ($\operatorname{Acc}@1$~\eqref{eq:ties}) is the expected accuracy with random tie-breaking. Unique top-1 counts cases where the correct value alone scores highest. Each value type has 16--17 cases. Overall: $\operatorname{Acc}@1=97.5\%$, unique top-1 $=95\%$, random baseline $=1\%$. SSN: Social Security number, MRN: medical record number.}
\label{fig:privacy}
\end{figure}

\noindent\textbf{Identification results.}
The correct value is the unique top-scoring candidate in 95 of the 100 cases. In the remaining five, it ties with one alternative, giving expected top-$1$ acc.\ $\operatorname{Acc}@1=97.5\%$~\eqref{eq:ties}. The five ties arise when two candidate values produce identical token multisets after tokenization, making them indistinguishable under the recovered-multiset scoring rule. When the program is allowed to return five or ten candidates, the correct value appears in every case (100\% top-5 and top-10 accuracy, Figure~\ref{fig:privacy}).

Accuracy is consistently high across all six value types (16--17 cases each), with no type falling below 88.235\%. This uniformity indicates that the attack is not limited to a particular format: it works equally well on short alphanumeric strings (passwords), structured numeric sequences (card numbers, SSNs and MRNs), and textual values (personal names and postal addresses).

\noindent\textbf{Control experiments.}
Two controls confirm that identification depends on the recovered private tokens, not on side information. First, without the recovered token identities, using only the known surrounding text or the cache token count leaves all 100 candidates equally plausible ($\operatorname{Acc}@1=1\%$, chance level). Second, when the requests are protected with different secrets while the attack retains its old estimates, identification drops to $\operatorname{Acc}@1=0.991\%$, again indistinguishable from guessing. The contrast (97.5\% vs.\ ${\sim}1\%$) confirms that the recovered private token content is the sole source of the identification signal.

\noindent\textbf{Implications.}
This experiment demonstrates that the privacy threat from token-multiset recovery extends beyond abstract text similarity to the disclosure of specific, actionable secrets. An infrastructure operator with cache access could, given a known request template and a candidate list, automatically identify which password, card number, or medical identifier a user submitted. The fully automated nature of the attack, requiring no human inspection of the reconstructed text, makes it scalable to large volumes of cached requests.

\begin{answerBox}
\textbf{Answer to RQ3.} Recovered tokens identify the correct sensitive value among 100 candidates with $\operatorname{Acc}@1=97.5\%$, completing the attack chain from orbit collapse through token recovery to the disclosure of specific private information.
\end{answerBox}

\subsection{RQ4: Defense Practicality}
\label{sec:defense}
The preceding results show that shuffled obfuscation leaks token content (RQ1--RQ2) and enables identification of specific sensitive values (RQ3), motivating a defense that prevents stored caches from exposing these tokens. We propose \veilkv{}, a byte-level authenticated-encryption scheme for KV caches, and evaluate it along three axes: correctness (whether protected caches preserve inference), security (whether scope and integrity checks reject unauthorized or tampered objects), and cost (the runtime overhead relative to the existing obfuscation).

\subsubsection{\veilkv{} Design and Integrity}
\label{sec:veilkv}
\begin{figure}[t]
\centering
\includegraphics[width=\columnwidth]{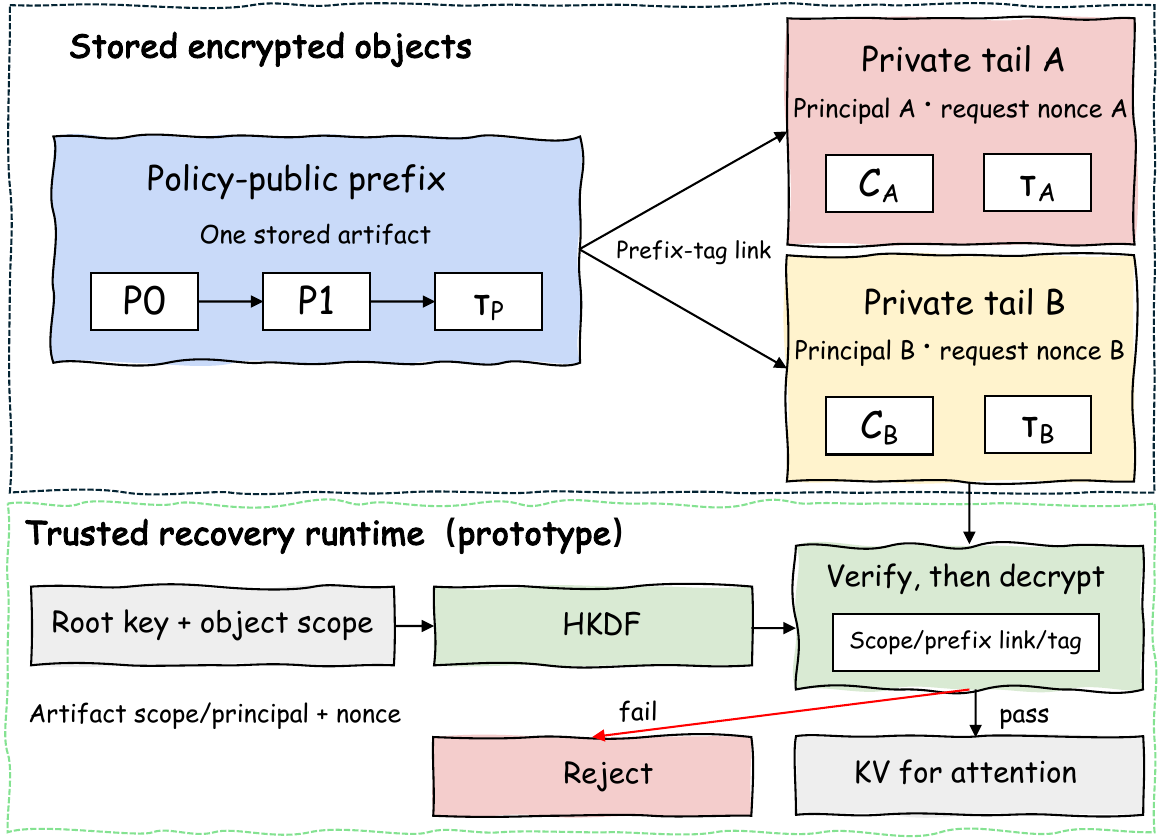}
\caption{\veilkv{} storage and recovery. A public prefix that the sharing policy permits requests to share is stored once. Each request's private tail is linked to this prefix and bound to its owner, request nonce and epoch. The trusted prototype checks these identities, the prefix linkage and authentication before releasing KV values.}
\label{fig:veilkv-sharing}
\end{figure}

\veilkv{} encrypts stored K/V bytes and authenticates them before use. Requests sharing a common public prefix store it once, while each request's private tail is bound to its owner (principal), request nonce and epoch (key version) (Figure~\ref{fig:veilkv-sharing}). Before releasing any KV values, the trusted adapter verifies these bindings and the authentication tag, then restores the datatype expected by the model (Appendix~\ref{app:veilkv-protocol}).

\noindent\textbf{Legitimate cache reuse.}
We verify that four authorized operations preserve the model's state and continuation, using independent unprotected paths as references: re-protecting a recovered cache, changing its owner or key version, protecting mid-generation, and reusing a stored public prefix with a private tail. Tests use 602 WildChat prompts on Qwen2.5-0.5B and Llama-3.2-1B in both FP32 and bf16. Because batched inference may pad requests with masked placeholder tokens, we test five padding layouts (none, or 8/16 tokens before or after the private text). All checks pass. Qwen3-14B and Qwen3-32B (bf16) also continue generation correctly from protected-and-restored caches (Appendix~\ref{app:defense602}).

\noindent\textbf{Scope and integrity enforcement.}
We mutate individual metadata fields (model, owner, request identifier), modify ciphertext bytes, and truncate stored objects. All 493{,}640 tamper checks reject the invalid input before any recovered state reaches the model. Replaying a previously used request nonce is likewise rejected, even after a service restart.

\noindent\textbf{End-to-end engine execution.}
We integrate \veilkv{} into vLLM 0.8.5.post1 on Qwen2.5-0.5B and Llama-3.2-1B (block sizes 16 and 32). The engine consumes restored KV pages during decoding. Tests cover ordinary requests, public-prefix reuse, private isolation and epoch transitions. For each request, the model generates six continuation tokens from a protected-and-restored cache. Prediction scores and token sequences match the unprotected reference exactly, confirming correct generation. All tamper cases are rejected before model execution (Appendix~\ref{app:vllm-live}). A separate reproducibility check shows that the unprotected engine itself exhibits minor numerical variability when bf16 prefix caching is enabled, establishing the baseline precision for our comparisons.

\subsubsection{Performance Comparison}
\label{sec:defense-comparison}
We compare four methods: unfused and fused \scheme{} (existing obfuscation baseline), authenticated encryption without prefix sharing (entire cache treated as private), and \veilkv{} (encryption with prefix sharing). The primary workload uses Llama-3.1-8B with FP32 caches. We sample real prompts from the main collection across five input-length groups (32--63 to 512--767 tokens) and time each operation 1{,}000 times per group on the same inputs.

At block size 16, \veilkv{}'s median protection time ranges from 103\,ms (shortest group) to 555\,ms (longest), and recovery from 37\,ms to 494\,ms. By contrast, unfused \scheme{} takes 32--33\,ms for protection and 50--52\,ms for recovery across all groups (Figure~\ref{fig:cache-latency-distribution}). The gap reflects the cost of trusted CPU cryptography and GPU--CPU transfers in byte protection, whereas \scheme{} operates entirely on the GPU. \veilkv{} is therefore functionally correct and supports prefix sharing, but at non-negligible additional latency compared to the obfuscation baseline (Appendix~\ref{app:veilkv-measurements}).

\begin{answerBox}
\textbf{Answer to RQ4.} \veilkv{} preserves correct inference and enforces scope and integrity checks in both native and vLLM execution, including public-prefix sharing. Its latency overhead relative to GPU-only obfuscation is non-negligible, reflecting the cost of byte-level cryptography and CPU--GPU transfers.
\end{answerBox}

\section{Discussion}
\label{sec:discussion}

\subsection{Orbit Collapse as a Paradigm Problem}
The attack on \scheme{} succeeds because a repeated-token input collapses the effective randomization from $b!$ permutations to $b$ distinguishable states. This collapse is not specific to \scheme{} but inherent to the shuffled obfuscation paradigm. The root cause is that security depends not on how many transformations the defender can sample, but on how many produce distinguishable outputs on the attacker's chosen input. For a repeated-token block, permutations that exchange identical rows are indistinguishable, leaving only $b$ observable states (one per marker position). The orbit size, not the group order, bounds the attacker's calibration difficulty. This reasoning applies to any defense in the paradigm. A designer must answer: what can an attacker distinguish after choosing the input? Counting all possible transformations overstates protection when many act identically on the chosen input. Additionally, auxiliary structures added for legitimate recovery (such as \scheme{}'s marker) may label the remaining states and reveal the reused secrets. Different marker and mixing designs require their own analysis, but the relevant security measure remains the observable orbit under adversarial inputs.

Section~\ref{sec:rq1} confirms this prediction empirically on two additional defenses~\cite{hidden,permllm}: both reduce to exactly $b$ distinguishable row arrangements under repeated inputs. Notably, although the three constructions use different mixing strategies (\scheme{} applies a secret linear transform and marker, Thomas et al.\ use sequence-level permutation alone, and PermLLM adds intra-row coordinate shuffling with a three-party LayerNorm), all converge to the same $b$-state orbit because the row-symmetry argument depends only on the input having identical rows and is agnostic to per-row transformations. The experiments (Section~\ref{sec:rq2}) further show that recovered parameters transfer across requests within the same secret epoch, enabling successful recovery across all tested models and configurations after a single calibration, including from caches produced during live vLLM inference.

\subsection{Defense Design Principles}
The attack exposes three principles that apply beyond \scheme{} to the shuffled obfuscation paradigm.

\noindent\textbf{P1: Do not reuse transforms across requests.} A refreshed permutation may vanish under a chosen input, leaving the reused transform exposed. Any defense that relies on permutation freshness while retaining other structural secrets faces this risk. The $S,a$ refresh experiment (Appendix~\ref{app:sa-continuation}) confirms this: after $S$ and $a$ change, the attacker's previous estimates no longer recover tokens accurately.

\noindent\textbf{P2: Bind secrets to the smallest practical scope.} Each private cache object should be bound to a specific owner, request nonce and epoch. When requests share a public prefix or continue across turns, the design must specify which secrets and metadata apply to every stored part.

\noindent\textbf{P3: Evaluate correctness and security jointly.} A modification that blocks the current attack is not a valid defense if it also prevents legitimate cache restoration. Conversely, failure of the attack on a modified design does not by itself establish security. \veilkv{} illustrates all three principles: it encrypts and authenticates cache bytes per request (P1--P2) and validates both byte-exact recovery and tamper rejection (P3).

\subsection{Threats to Validity}
\label{sec:threats}

\noindent\textbf{Construct validity.}
We use complementary metrics at each stage of the attack chain: bag recall for token-content recovery, position acc.\ and exact match for ordering quality, embedding cosine for semantic similarity, and candidate $F_1$ with top-$k$ acc.\ for sensitive-value identification. Together, these capture content, order and privacy impact. The sensitive-value experiment uses a closed candidate set with known surrounding text, which scopes the identification claim to this controlled setting.

\noindent\textbf{Internal validity.}
To ensure results are not artifacts of a particular secret key, we repeat recovery under five independently generated configurations and keep prompts fixed when testing secret changes. For cache correctness, protected and unprotected paths receive identical tokens, masks and positions, isolating protection-induced differences. Bootstrap resampling keeps all observations from one prompt together across heads, seeds and configurations.

\noindent\textbf{External validity.}
The target sample covers English WildChat conversations across five length groups. Recovery targets first-layer Value blocks because dictionary lookup requires position-independent token representations, which holds at the first layer (Section~\ref{sec:prelim}). The attack requires the probe and target to share the same $S$, $M$ and $a$, which holds within a secret epoch of the tested implementation. The orbit-collapse mechanism is confirmed in two additional defenses~\cite{hidden,permllm}, though each construction would need its own parameter-recovery method. \veilkv{} is validated in the tested software and vLLM versions. A production deployment would additionally need hardware-backed key isolation and anti-rollback guarantees for the epoch ledger.

\section{Related Work}
\label{sec:related}

\noindent\textbf{Input symmetry and secret reuse.}
Classical permutation-cipher attacks exploit chosen inputs to reduce the effective key space~\cite{permutationciphers}. In the LLM setting, GELO identifies repetition and Gram leakage under fresh per-batch mixing~\cite{gelo}, and ConjFormer enables inference under orthogonal conjugation~\cite{conjformer}. \scheme{} itself recognizes repetition-induced permutation collapse and introduces a recovery marker to support legitimate decoding~\cite{kvcloak}. Our contribution is to show that this marker still limits the orbit to $b$ states under repeated text, and that rank-one differences between these states algebraically identify the reused transforms. Unlike prior observations that note symmetry as a concern, we convert the reduced orbit into a complete, reusable demixer that transfers across requests within the same secret epoch.

\noindent\textbf{Relation to Thomas et al.~\cite{hidden}.}
Thomas et al.\ reconstruct tokens from row-, coordinate- and factorized-permuted hidden states through prefix-conditioned candidate matching, using sorted $L_1$ distance for coordinate permutations. Their attack targets permutation-only defenses without dense mixing. Our attack addresses the additional row transform $S$, column transform $M$ and additive marker in \scheme{}, which are absent from the defenses Thomas et al.\ evaluate. We first derive an equivalent demixer from protected probe states and then recover first-layer Value-block multisets with a public-model dictionary. This algebraic calibration is reusable across requests, whereas Thomas et al.'s candidate matching operates per prefix and per block.

\noindent\textbf{Other structural attacks on inference protection.}
Several concurrent works exploit reused structural secrets in non-permutation defenses. Saini et al.\ identify reused noise subspaces with zero and basis-vector queries~\cite{partialtee}. Collapse aligns obfuscated weight views and separates masks~\cite{collapse}. ArrowRevelio, SpectralLeak/LatticeLeak and BiShield-TEE study shared directions or unilateral weight transforms~\cite{arrowrevelio,spectralleak,bishield}. These attacks typically require access to model weights or intermediate activations. Our method instead operates on stored cache blocks only: input symmetry suppresses the effect of fresh shuffling, and marker-state differences expose the remaining transforms without accessing the defender's transformed model weights.

\noindent\textbf{Cache protection and authorized reuse.}
KV cache sharing accelerates serving~\cite{paged}, but creates privacy risks: EarlyBird exploits timing side channels from shared prefixes~\cite{earlybird}, and PromptPeek~\cite{promptpeek2025} recovers prompt content from cache metadata. SafeKV, CachePrune and CacheSolidarity selectively share or isolate caches to mitigate reuse side channels~\cite{safekv,cacheprune,prefixwall}. These defenses address access-pattern or timing leakage but do not protect the stored bytes themselves. \veilkv{} instead encrypts and authenticates cache bytes, binding reusable public prefixes to private continuations scoped by owner, request and epoch, and checking these associations before any KV values are released to the model. Confidential-inference designs protect computation through different trust boundaries~\cite{sok}: CryptoGen combines homomorphic encryption and secret sharing~\cite{cryptogen}, Cachemir optimizes FHE-based KV-cache inference~\cite{cachemir}, and Cascade distributes token shards~\cite{hidden}. \veilkv{} occupies a different point in this design space: it uses standard key derivation and encrypt-then-MAC~\cite{hkdf,etm} for stored bytes rather than computation-level encryption, restores native tensors within a trusted boundary, and validates authorized sharing through physical-page consumption in vLLM.

\noindent\textbf{Text recovery and privacy measurement.}
Embedding-inversion studies reconstruct text from representations using learned mappings~\cite{morris2023,li2023,dong2025,deepinvert} or co-occurrence alignment~\cite{pma}. These methods typically operate on dense representations (sentence embeddings, hidden states) and produce approximate reconstructions. Our recovery is instead anchored by exact algebraic equations: the demixer yields token identities with 100\% multiset accuracy, and a separate language-model ordering step~\cite{wordorder,ibis} infers within-block positions. We complement content recovery with two additional privacy measurements, semantic similarity and sensitive-value identification, to demonstrate that the recovered tokens constitute an actionable threat.

\section{Conclusion}
We showed that input symmetry collapses the $b!$ nominal randomization of shuffled KV cache obfuscation to as few as $b$ observable states, enabling the Repeat-to-Reveal attack. On \scheme{}, a single repeated-token calibration recovers an equivalent demixer that achieves 100\% token-multiset recovery across eleven models (0.5B--32B parameters), thirteen configurations, with additional checks on vLLM inference engine, reconstructs text at $S_{\mathrm{cos}}=0.947$ semantic similarity, and identifies sensitive values at 97.5\% top-1 accuracy. Checks on two additional defenses confirm the same reduction in observable row structure. These results establish a paradigm-level vulnerability in shuffled obfuscation and motivate defenses based on authenticated encryption rather than algebraic obscurity. As a first step, \veilkv{} replaces obfuscation with byte-level authenticated encryption supporting scoped prefix sharing.

\clearpage
\bibliographystyle{IEEEtran}
\bibliography{references}
\appendices
\raggedbottom
\setcounter{dbltopnumber}{2}
\section{Ethics Considerations}
\label{app:ethics}

\noindent\textbf{Data and privacy.}
Experiments use local public checkpoints, synthetic text and public WildChat-4.8M conversations under the ODC-By license~\cite{wildchat48m}. We select English conversations whose toxic and redacted flags are false. These filters do not guarantee that the text contains no sensitive information. The retained experimental records include user prompts and model and timestamp metadata, but exclude identifiers derived from IP addresses, request headers and source assistant responses. The sensitive value examples are synthetic. API ordering uses token collections from the public dataset (Appendix~\ref{app:api-order}).

\noindent\textbf{Responsible disclosure.}
After contact via GitHub and email, the \scheme{} authors enabled private vulnerability reporting and invited a detailed report. We submitted it on November 3, 2026, and later added technical details. The authors confirmed the report and acknowledged the vulnerability on November 17, 2026. They also expressed willingness to coordinate disclosure timing.

\section{Reproducibility and Defensive Checks}
\label{app:repro}

\subsection{Implementation and Software}
\label{app:implementation-pointers}
The official \path{get_kvcloak_config.py} and \path{KVCloak} define the tested transforms and when their secrets are reused. Section~\ref{sec:methodology} specifies the model interfaces and execution environment.

\subsection{Selection of the 602-Prompt Subset}
\label{app:subset602}
We assign each main prompt to the first matching category using text rules, in this order: code (code fences or programming keywords), mathematics (LaTeX notation or terms such as ``solve,'' ``integral'' and ``probability''), creative writing (writing terms in texts longer than 500 characters), question answering (question forms in the first 300 characters of texts shorter than 800 characters), and general text (the remainder). These are coarse heuristic labels. Only 52 of the 3,000 prompts match the mathematics rules after code is assigned first, so we retain all 52. We sample the other categories with nearly equal quotas, allocating within each category approximately in proportion to the available length groups and shuffling with seed 42. Integer rounding yields 602 prompts: 137 code, 52 mathematics, 138 creative writing, 137 question answering and 138 general text.

\subsection{\veilkv{} Protocol}
\label{app:veilkv-protocol}
\veilkv{} uses AES-128-CTR and authenticates the ciphertext after encryption~\cite{etm}. HKDF~\cite{hkdf} derives separate material for encryption, authentication and initialization vectors. The trusted component verifies authentication before decrypting the cache and restoring the datatype required by the model. For a shared public prefix, the authenticated identity includes the model, tokenizer, RoPE configuration, token prefix, layer, K/V role, block, datatype and sharing policy. A private tail is additionally bound to its owner and request nonce. The authenticated metadata also records the cache coordinates, epoch, policy, sequence length and the preceding authentication tag. The final tag of a stored prefix starts the authentication chain for its private tail. A persistent SQLite ledger rejects reuse of a request nonce and conflicting versions of a public prefix. The prototype simulates hardware controlled key release and assumes that trusted storage prevents rollback to an older ledger state.

\subsection{Matched Control}
\label{app:matched-controls}
The secret replacement test keeps the same 602 Qwen2.5-0.5B requests and both KV heads in all three conditions. Recovery counts before replacement, after replacing $S,M,a$ while retaining the old attack estimates, and after a new calibration are $359{,}936/359{,}936$, $1{,}482/359{,}936$ and $359{,}936/359{,}936$, respectively.

\subsection{\veilkv{} Measurements}
\label{app:veilkv-measurements}
We group the main collection by the number of tokens produced by each tested model's tokenizer: 32--63, 64--127, 128--255, 256--511 and 512--767. Within each group, we sample complete prompts uniformly with replacement and make 1,000 timed calls for each operation. These calls cover 475--497 distinct prompts because a prompt may be sampled more than once. All methods and block sizes for one model use the same sampled prompts. We test Llama-3.1-8B in FP32 with block sizes 16, 32 and 64, and Qwen2.5-0.5B in bf16 with block size 16. For every profile and length group, we time protection and recovery for four methods and time native prefill separately. This produces 180,000 formal timings. Each series also has 20 warmup calls. All measurements use the A100 and 16 host threads.

\begin{figure*}[t]
\centering
\includegraphics[width=0.76\textwidth]{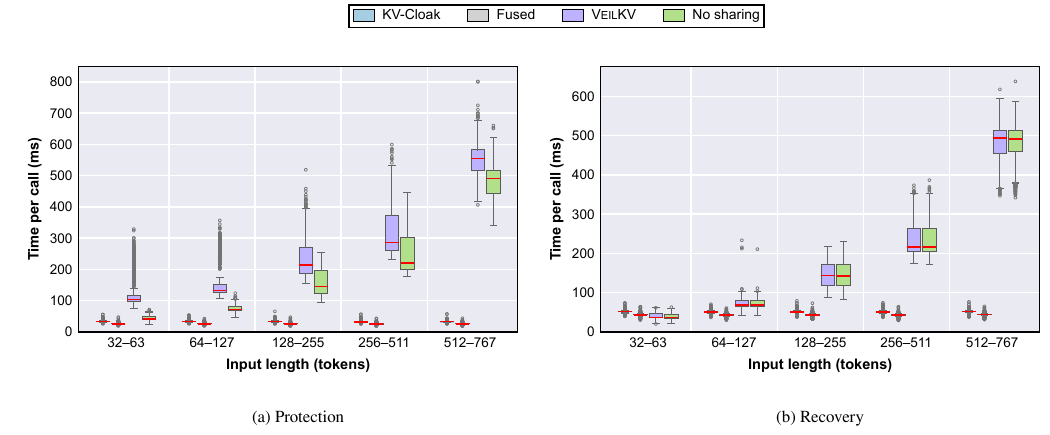}
\caption{Time for cache operations by input length on Llama-3.1-8B, using FP32 caches, 16 host threads and KV-Cloak block size 16. Each combination of method, operation and length group contains 1,000 timed calls on the same sampled prompts. Boxes span the middle 50\% of timings, and red lines mark medians. Whiskers extend to the furthest samples within 1.5 interquartile ranges, and all remaining samples are shown as points. Times are milliseconds per call. Byte recovery includes restoration of the data type required by the model. Fused denotes fused KV-Cloak, and No sharing denotes authenticated encryption without prefix sharing.}
\label{fig:cache-latency-distribution}
\end{figure*}

Figure~\ref{fig:cache-latency-distribution} shows the Llama profile with block size 16. For each sampled prompt, we randomize the order of the method and operation calls and synchronize CUDA immediately before and after each timed call. Timings for byte protection include cryptography on the CPU and transfers between CPU and GPU memory, while KV-Cloak keeps its tensors on the GPU. The sharing profile treats the first 32 tokens as public, while the baseline without sharing treats the entire request as private. Before timing begins, we construct the cache by processing the public prefix and then the private text. Preparation of fused inputs and protection of the input used by a recovery call also occur before timing. Native prefill is timed as a separate operation. Normalized values use the actual token count and cache size of each call. We report each model, datatype, block size and length group separately.

\subsection{Refreshing Row Secrets and Continuing Inference}
\label{app:sa-continuation}
\noindent\textbf{Attack transfer.}
We generate new values for $S$ and $a$ while keeping the column transform $M$ fixed. On the same 602 Qwen2.5-0.5B prompts and both KV heads, recovery using the attacker's old estimates falls from 100\% to $6{,}288/359{,}936$ (1.747\%).

\noindent\textbf{Legitimate continuation.}
We test the 602 prompts with two models and five ways of placing masked padding, giving 6,020 input fixtures. Testing both FP32 and bf16 gives 12,040 pairs that compare the original and refreshed secrets. Each fixture contains a 16 token prefix, no more than 134 body tokens and three continuation segments containing 2, 3 and 1 tokens. The two secret conditions use the same attention masks, token positions and permutations. A segment passes when its outputs are finite and have the expected shape, its most likely token matches the unprotected reference, and its largest logit difference is at most 0.001 in FP32 or 0.6 in bf16. In FP32, the original and refreshed secrets pass 11,780/18,060 and 11,764/18,060 segments. In bf16, they pass 172/18,060 and 183/18,060. The totals are therefore 11,952/36,120 and 11,947/36,120. Among the 36,120 paired segments, 206 pass only with the original secrets and 201 pass only after refresh. In FP32, all 12,336 checks whose 16 position block contains both masked and unmasked positions fail. The other checks pass 23,544/23,784 times. Two independent unprotected runs agree exactly. The largest difference between full and cached FP32 inference is $3.819\times10^{-4}$, and the most likely token remains the same.

\subsection{Candidate Control for Numerical Masking}
\label{sec:arithmetic-profile}
We also test whether simpler numerical summaries can identify a target without the attack's calibration. This control uses a variant with secret rotations of rows and columns and with row scaling. For Qwen and Llama, the first 16 token block from each of 602 WildChat prompts yields 291 and 283 distinct token sequences, respectively. Every target is compared with 99 random token sequences. Matching the eigenvalues of $VV^{\T}$ identifies 258/602 Qwen targets and 402/602 Llama targets. Uniform guessing would identify 1\%. Comparing sorted row norms identifies 30/602 and 176/602. A Procrustes transformation~\cite{procrustes} fitted from public sentences identifies 75/602 and 13/602.

\setcounter{dbltopnumber}{3}

\begin{figure*}[t]
\centering
\includegraphics[width=0.72\textwidth]{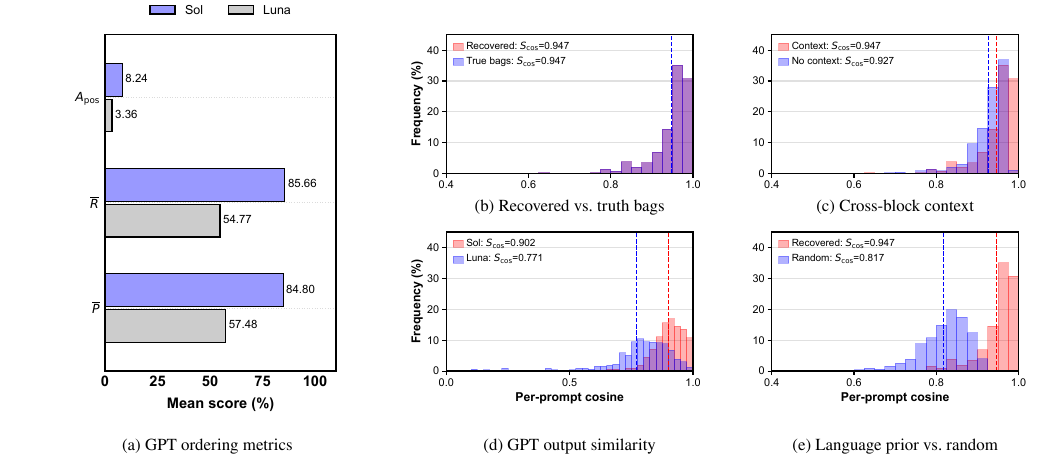}
\caption{Ordering on the same 602 prompts. (a) GPT position accuracy $A_{\mathrm{pos}}$, average recall $\overline R$ and precision $\overline P$. (b--e) Each histogram shows the percentage of prompts in each embedding cosine range, retaining all 602 scores. Dashed lines and legend values give the mean $S_{\mathrm{cos}}$. True bags contain the original tokens and counts without their order. Local search receives tokens grouped by cache block. GPT receives one combined token list for each prompt.}
\label{fig:ordering-distributions}
\end{figure*}

\section{Measurement Denominators and Algebra}
\label{app:metrics}
For configuration $c$, let $H_c$ be the head count, $b_c$ the block size, and $L_{c,q}$ the tokenized length of prompt $q$. With $Q$ prompts and two permutation seeds,
\begin{equation}
N_c=2H_c\sum_{q=1}^{Q} b_c\lfloor L_{c,q}/b_c\rfloor.
\label{eq:denominator}
\end{equation}
Calibration succeeds for every head in the main study. We compute the recovery interval from 2,000 bootstrap samples. Each bootstrap sample draws whole prompts within each length group, and keeps all configurations, heads and seeds belonging to one prompt together. With seed 20261029, the interval is [100\%, 100\%].

\subsection{Algebraic Identities}
\label{app:bounds}
For multiplicities $n_i$ of all distinct rows of $A$, including zero rows, $|\mathrm{Stab}(A)|=\prod_i n_i!$. With $w'=-w-2a/b$ and $q_n(k)$ the probability of seeing $k$ states after $n$ blocks,
\begin{align}
\|w'\|^2-\|w\|^2&=\frac4b\langle w,a\rangle+\frac4{b^2}\|a\|^2,\\
q_{n+1}(k)&=\frac kb q_n(k)+\frac{b-k+1}{b}q_n(k-1).
\end{align}
Here $q_0(0)=1$, and probabilities for impossible values of $k$ are zero. Figure~\ref{fig:probe} uses $p_1=q_n(b)$. The probability that both heads obtain complete coverage lies between $\max(0,2p_1-1)$ and $p_1$, without assuming that the two heads are independent. These bounds describe state coverage rather than whether the decoder succeeds.

\section{Reconstruction and Identification Protocols}
\label{app:ordering}
The ordering study uses 602 of the 3,000 main prompts. The same prompts are supplied to all four local conditions and both GPT models. From shortest to longest, the five length groups contribute 101, 112, 119, 104 and 166 prompts. We evaluate complete blocks of 16 tokens and the recovered token multisets from the first KV head, head~0. After selecting the complete blocks, the 602 prompts contain 574 distinct token sequences.

\subsection{API Ordering}
\label{app:api-order}
GPT receives one combined list containing the token IDs, decoded token pieces and repetition counts for an entire prompt. The local beam instead receives a separate token multiset for each cache block and must use every supplied token exactly as many times as it appears. Sol uses medium reasoning and Luna uses high reasoning. Both use temperature zero and a limit of 4,096 completion tokens, and return only reconstructed text. We retain all 602 outputs from each model, including empty responses and responses that stop after reaching the completion limit. Sol and Luna exactly reconstruct 4/602 and 0/602 target sequences. They preserve all target tokens and their counts in 47/602 and 0/602 outputs. For each prompt, recall divides matching token occurrences by the target length, and precision divides them by the output length. Empty output receives zero precision. Figure~\ref{fig:ordering-distributions} reports their per-prompt means, $\overline R$ and $\overline P$. We embed every target and output independently with \textnormal{all-MiniLM-L6-v2}, truncating each input at 256 encoder tokens. Figure~\ref{fig:ordering-distributions} retains every cosine score, and an empty output receives a score of zero.

\section{Numerical Diagnostics and Extended Evaluation}
\label{app:saved-record-review}
Rare algebraic ambiguities can be addressed by choosing a different repeated token for the probe. Token recovery also requires the public model's Value vectors to distinguish candidate tokens. Before decoding targets, the implementation verifies that Value vectors computed from the public checkpoint agree with the corresponding states observed on the attacker's own probe.

\subsection{Calibration Costs}
\label{app:cpu-costs}
We repeat the complete calibration session five times for each model. The median times are 0.853 seconds for Qwen and 1.847 seconds for Llama. A session includes probe construction, model prefill, cache protection and calibration on the CPU. It excludes model loading, threshold estimation and construction of the public token dictionary. Separate saved CPU measurements show that target decoding takes 43.4 seconds per 1,000 target tokens for Qwen and 120.0 seconds for Llama. These decoding measurements include diagnostic work and varying contention from other processes on the host.

\subsection{Scaling Beyond the Default Configuration}
\label{app:scaling-followup}

\subsection{Expanded Validation of Protected Cache Bytes}
\label{app:defense602}
The 602 prompts are tested with two models and five ways of placing masked padding, giving 6,020 input fixtures. Testing both datatypes gives 12,040 cases for each operation. We test protecting a recovered cache again, changing its authorized identity and epoch, protecting it after generation has begun, and reusing a stored public prefix with private text. Each operation passes 12,040/12,040 requests, for 48,160 passes in total. Repeated protection covers epochs 3--5 while the model generates continuation segments of 2, 3 and 1 tokens. Two independent unprotected reference paths provide 96,320 aligned comparisons with the model consuming the cache. Every recovered K/V tensor has the expected datatype and exactly the same bytes as its reference.

All 493,640 attempts to use the wrong model, owner, request, epoch or cache coordinates, to modify ciphertext, or to truncate a stored object are rejected. We also test the persistent nonce ledger with 3,200 competing operations across 100 rounds and eight processes. The ledger rejects replay after a service restart, and 1,000 resource checks leave all four database handles in their expected state.

\subsection{Attack and Defense in vLLM}
\label{app:vllm-live}
The vLLM 0.8.5.post1 integration uses the V0 engine with one process, FlashAttention, eager execution and bf16 caches. Stored K/V data are removed from the referenced GPU pages between calls. Before the model uses those pages again, the trusted component authenticates and restores the cache. Requests with the same declared public prefix share its first 32 tokens and their physical pages. The owner and request identifiers keep their private pages separate before scheduling. In 40 functional cases, the epoch changes after the model generates two tokens.

We test Qwen2.5-0.5B and Llama-3.2-1B with physical block sizes 16 and 32, using 602 fixed WildChat inputs for each model. Every combination of model and block size contains 602 ordinary requests, 301 pairs that share a public prefix, 602 pairs that use the same private text under different identities, and 4,816 attempts to modify or misuse protected data. Two independent unprotected references generate six continuation tokens for comparison.

\noindent\textbf{Reproducibility of the engine reference.}
A separate check repeats unprotected generation for all 602 Qwen2.5-0.5B inputs. With bf16 prefix caching, two runs agree on 555/602 inputs at block size 16 and 539/602 at block size 32. Disabling prefix caching gives 602/602 agreement. The selected divergences have score gaps of 0--0.125 between the leading candidates, and FP32 removes the differences on the diagnostic inputs. Because the reference itself is numerically sensitive, the \veilkv{} tests retain exact paired comparisons with an independently executed reference.

\subsection{Orbit Checks in Public Permutation Implementations}
\label{app:extra-orbit}
Both CPU studies use 120 fixed probes at the first layer of Llama-3.2-1B. For Thomas et al.'s implementation, we draw 1,024 sequence permutations at block sizes 4, 8, 16 and 32. A control using fresh coordinate permutations retains 128/128 distinct outputs. PermLLM retains 512/512 distinct raw outputs before values within each row are sorted. At block size 4, controls with increasing repetition produce 4, 12 and 24 states. All 1,280 checks of PermLLM's native protocol remain within the noise measured on a fixed input, and its LayerNorm error stays below $7.48\times10^{-5}$.

\end{document}